\documentclass[journal,comsoc]{IEEEtran}
\IEEEoverridecommandlockouts

\usepackage{url}
\usepackage{times}
\usepackage{amsmath}
\usepackage{epsfig}
\usepackage{multirow}
\usepackage{longtable}
\usepackage{amsfonts}
\usepackage{amssymb}%
\usepackage{color}
\usepackage{bm}
\usepackage{epstopdf}
\usepackage{subfigure}
\usepackage{cite}
\usepackage{graphicx}
\usepackage{amsmath}
\usepackage{times}
\usepackage{latexsym}
\usepackage[center]{caption2}
\usepackage{stfloats}
\usepackage{cases}
\usepackage{array}
\usepackage{setspace}
\usepackage{fancyhdr}
\usepackage{balance}
\usepackage{algorithm} 
\usepackage{algpseudocode} 

\usepackage[table]{xcolor}
\usepackage{makecell}

        {\par\noindent{\bf Proof.}}%
        {{\hfill{\large$\Box$}}\smallskip}

\begin{document}

\title{
Responsive Regulation of Dynamic UAV Communication Networks Based on Deep Reinforcement Learning}

\author{\IEEEauthorblockN{
Ran~Zhang,~\IEEEmembership{Member,~IEEE,}
Duc~Minh~(Aaron)~Nguyen,
Miao~Wang,~\IEEEmembership{Member,~IEEE,}
Lin~X.~Cai,~\IEEEmembership{Senior~Member,~IEEE,}
and~Xuemin~(Sherman)~Shen,~\IEEEmembership{Fellow,~IEEE}}
\thanks{R. Zhang, M. Nguyen, and M. Wang are with the Department
of Electrical and Computer Engineering, Miami University, Oxford, OH, 45056 USA, email: \{zhangr43,nguyendm,wangm64\}@miamioh.edu. L. X. Cai is with Department of Electrical and Computer Engineering, Illinois Institute of Technology, Chicago, IL, 60616 USA, email: lincai@iit.edu. X. Shen is with Department of Electrical and Computer Engineering, University of Waterloo, Waterloo, ON, N2L 3G1 Canada, email: sshen@uwaterloo.ca.}
\thanks{The research is partially supported by the National Science Foundation under grant ECCS-1554576.}
}


\maketitle

\IEEEpubidadjcol

\begin{abstract}
In this chapter, the regulation of Unmanned Aerial Vehicle (UAV) communication network is investigated in the presence of dynamic changes in the UAV lineup and user distribution. We target an optimal UAV control policy which is capable of identifying the upcoming change in the UAV lineup (quit or join-in) or user distribution, and proactively relocating the UAVs ahead of the change rather than passively dispatching the UAVs after the change. Specifically, a deep reinforcement learning (DRL)-based UAV control framework is developed to maximize the accumulated user satisfaction (US) score for a given time horizon which is able to handle the change in both the UAV lineup and user distribution. The framework accommodates the changed dimension of the state-action space before and after the UAV lineup change by deliberate state transition design. In addition, to handle the continuous state and action space, deep deterministic policy gradient (DDPG) algorithm, which is an actor-critic based DRL, is exploited. Furthermore, to promote the learning exploration around the timing of the change, the original DDPG is adapted into an asynchronous parallel computing (APC) structure which leads to a better training performance in both the critic and actor networks. Finally, extensive simulations are conducted to validate the convergence of the proposed learning approach, and demonstrate its capability in jointly handling the dynamics in UAV lineup and user distribution as well as its superiority over a passive reaction method.
\end{abstract}


\section{Introduction}\label{sec.Intro}
Unmanned aerial vehicles (UAVs) have been attracting increasing attention as a key component in the future wireless communications\cite{zhang2019iot}. Compared to the terrestrial base stations (BSs), UAVs equipped with wireless transceivers can serve as mobile BSs and stand out in providing highly on-demand services, better wireless connectivity to the ground users, and much lower deployment cost due to the almost infrastructure-free network construction\cite{zeng2016wireless}. As reported in \cite{market}, the UAV market is estimated at USD 27.4 billion in 2021 and is projected to reach USD 58.4 billion by 2026. In this booming market, UAVs have been exploited in many applications such as mobile edge computing\cite{li2020energy,shi2020mean}, crowd/traffic surveillance\cite{motlagh2017uav}, emergency rescue\cite{zhao2019disaster}, cached content delivery\cite{chen2017caching}, network coverage enhancement and extension\cite{nasir2019uav}, etc. 

Various aspects of UAV-based communications have been extensively studied, ranging from radio resource allocation and trajectory design, to energy management and computing offloading \cite{mozaffari2019tutorial,wu2019fundamental}. Conventional approaches typically formulate the studied problems into (mixed integer) non-convex optimization problems. The original NP-hard problem is generally decoupled into a set of sub-problems and solved by iterative algorithms\cite{zeng2018trajectory,wu2018common,zeng2019energy,li2020energy,guo2019uav,mozaffari2018beyond,wu2018joint,mozaffari2017mobile,yang2019energy}. This conventional methodology is a better fit where the network parameters are fixed. In UAV-based communications, parameters such as the network topology, wireless channel conditions and user distributions are usually time-varying due to the mobility of UAVs, topographic relief and the temporality of the on-demand services. As a result, the above methods need to be re-executed each time the parameters are updated. With the exponentially increasing network scale and heterogeneity in the future, it will be increasingly difficult for conventional approaches to handle the network dynamics.

Thanks to recent advances in machine learning\cite{alpaydin2020introduction}, reinforcement learning (RL) \cite{RL,luong2019applications} is becoming a promising solution to UAV communication problems. By constantly interacting with the environment and learning from the interaction experiences, RL agents are strongly capable of making sequential decisions in time-varying environments free of the environment models. Existing RL-based studies on UAV communications focus mainly on control policy development given a fixed set of UAVs\cite{klaine2018distributed,cui2019multi,hu2020reinforcement,liu2019trajectory,hu2020distributed,singh2018distributed,challita2019interference,tang2020deep,cheng2019space,liu2019optimized,liu2018energy,khairy2020constrained}. Few works have investigated how the network should be regulated considering the dynamic lineup change of the serving UAVs. Due to the ad hoc nature of the UAV networks, the serving UAV lineup can dynamically change at times. UAVs have to quit the network when their batteries are depleted; supplementary UAVs can also join the serving lineup whenever needed. Either case will inevitably create fluctuations in the network performance, thus calling for responsive regulation strategies when such changes happen. When regulated, the network should not passively react after the change, but identify the upcoming change and take actions in advance to minimize (or maximize) the performance loss (or gain) during the transition to the new optimal UAV positions. Such procedure is referred to as proactive self-regulation (PSR) in this paper. A major challenge of using RL for PSR include is that the dimensions of the state and action space both change during the training process, which is irregular for RL. Another challenge is how to promote the learning exploration around the time of change so that the agent is able to take actions in advance.
In addition, most of the existing works only consider stationary user distribution, whereas the distribution can be dynamic in practice. The works \cite{liu2019trajectory} and \cite{huang2020online} considered user mobility in the RL framework, but the UAV trajectories are limited to a mesh grid.


Motivated by the above considerations, PSR of a UAV communication network is investigated in this chapter with dynamic change in UAV lineup and user distributions. We aim to achieve an optimal UAV control policy via DRL which relocates the UAVs in advance when $i$) at least one UAV is about to quit or join the network, or $ii$) the user distribution changes, rather than passively relocates the UAVs after the change. To the best of our knowledge, this is the first work of optimal regulation of a UAV communication network that jointly considers the dynamic UAV lineup and user distribution. The contributions are given as follows.
\begin{itemize}
    \item A DRL-based approach for PSR of UAV communication network is developed. The approach aims to maximize the accumulated user satisfaction (US) score of the considered time horizon where the change in UAV lineup happens. 
    To accommodate the continuous state space and action space, the state-of-the-art actor critic learning method, DDPG\cite{lillicrap2015continuous}, is selected among all the DRL variants so that the UAV battery status, positions and movements can be accurately recorded.
    \item To promote the learning exploration around the lineup change and achieve better training performance on both the actor and critic networks, an asynchronous parallel computing (APC) structure is proposed. The proposed PSR approach under APC is referred to as PSR-APC. 
    \item The PSR-APC approach is further extended to the case of dynamic user distribution. Time is integrated as one of the learning states to achieve a time-dependent control policy.
    \item Extensive simulations are conducted to demonstrate the convergence and efficacy of the proposed PSR-APC approach. Compared to a passive reaction method, the proposed approach achieves surpassing accumulated US scores during the transition period.
\end{itemize}


The reminder of the paper is organized as follows. Section \ref{sec.SystemModel} describes the system model and formulates the problem. Section \ref{sec.PK} introduces preliminary knowledge on DRL and the adopted DDPG algorithm. Section \ref{sec.Algorithm} elaborates the detailed design of the proposed APC-PSR approach. Section \ref{sec.Dynamic} extends the APC-PSR from fixed user locations to dynamic user distributions. Numerical results are presented in Section \ref{sec.Simulation}. Finally, Section \ref{sec.Conclusion} concludes the paper.

\section{Related Works}\label{sec.Related}
The conventional optimization or rule-based methods have been extensively applied to UAV communications. For instance, Nasir \emph{et al.} \cite{nasir2019uav} and Zeng \emph{et al.} \cite{zeng2018trajectory} studied the resource allocation (RA) and trajectory design problem of a single UAV to maximize the minimum user rate and the mission completion time, respectively. When multiple UAVs are present, UAVs need to be coordinated in interference management, trajectory design, and user association. Mozaffari \emph{et al.} \cite{mozaffari2018beyond} studied the joint UAV positioning, frequency planning, and user association problem to minimize the UAV-user latency. Wu \emph{et al.} \cite{wu2018joint} additionally considered UAV trajectory design and power control. Mozaffari \emph{et al.} \cite{mozaffari2017mobile} focused on energy consumption and minimized the total propulsion energy of UAVs.  

When the network environment is time-varying and sequential decisions need to be made, RL-based UAV control approaches have been studied. Many existing works rely on a centralized agent to learn optimal joint policies for all the network entities. For instance, Singh \emph{et al.}\cite{singh2018distributed}, Challita \emph{et al.}\cite{challita2019interference} and Tang \emph{et al.}\cite{tang2020deep} applied deep Q-learning (QL) to optimize the RA, interference management, and trajectory design of UAVs, respectively, in UAV-assisted cellular networks. Liu \emph{et al.} \cite{liu2019optimized} employed double QL to design optimal UAV trajectories that maximize the number of satisfied users with time-constrained requirements. 
To accommodate large action space and expedite convergence, actor-critic (AC) based deep RL (DRL) is applied. Cheng \emph{et al.} \cite{cheng2019space} proposed an AC RL approach to optimize RA and task scheduling in UAV-assisted computing offloading. Khairy \emph{et al.}\cite{khairy2020constrained} studied the joint altitude control and channel access problem of a solar-powered UAV network by employing actor-critic RL. Liu \emph{et al.} \cite{liu2018energy} targeted the energy concerns of UAVs and exploited the up-to-date AC variant, i.e., deep deterministic policy gradient (DDPG) algorithm, to jointly maximize the energy efficiency, user fairness and network coverage. 


Multi-agent RL (MARL) has been exploited in a few existing works to make the learning distributed and scalable to the network size. For instance, Klaine \emph{et al.} \cite{klaine2018distributed} proposed a distributed QL approach to find the UAV positions that maximize the total amount of covered users. Cui \emph{et al.} \cite{cui2019multi} and Hu \emph{et al.} \cite{hu2020reinforcement} applied multi-agent QL to optimize RA and trajectory design, respectively. Liu \emph{et al.} \cite{liu2019trajectory} developed a multi-agent QL framework to optimize the UAV trajectory and power control, considering the ground user mobility. Hu \emph{et al.}\cite{hu2020distributed} proposed a value decomposition based MARL solution coupled with a meta-training mechanism to accelerate the learning of multi-UAV trajectories while generalizing the learning to unfamiliar environments. Pham \emph{et al.}\cite{pham2018cooperative} and Chen \emph{et al.}\cite{chen2020mean} integrated game theories into MARL to solve the complex dynamic of the joint UAV actions and simplify the complex interactions between multiple objectives and multiple UAVs, respectively. 

All the above works consider a fixed set of serving UAVs. The proposed work will be among the first to fill this research gap.

\section{System Model and Problem Formulation}\label{sec.SystemModel}
In this section, the system model is first depicted, followed by the problem formulation.

\subsection{Network Environment}\label{subsec.nt}
\begin{figure}[!ht]
	\centering
	\includegraphics[width=3.4in]{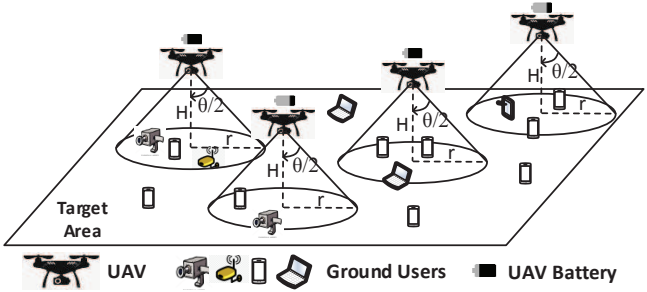}
	\caption{UAV coverage range as a disk area.} \label{fig.SystemModel}
\end{figure}

As illustrated in Fig. \ref{fig.SystemModel}, we consider a target area $\mathbf{A}$ with a set $\mathbf{S}_{ur}$ of $N_u$ ground users served by a lineup $\mathbf{S}_{UAV}$ of $N_{UAV}$ UAVs. The target area is an $L$-by-$L$ square. A large percent of the users are randomly distributed around several separate hot spots while the remaining are uniformly distributed throughout $\mathbf{A}$. The UAVs fly within $\mathbf{A}$ at a fixed altitude $H$ to serve the ground users with guaranteed minimum throughput. 
The antennas of each UAV are strongly directional such that the transmit power is concentrated within an aperture angle of $\theta$ right below the UAV. As a result, the coverage of a UAV on the ground is a disk area with radius $\textbf{r}=H\tan(\frac{\theta}{2})$, as shown in Fig. \ref{fig.SystemModel}. Users will not be interfered by one UAV if they are outside its coverage disk. 

\subsection{Spectrum Access}\label{subsec.sa}
All the UAVs are connected to external networks via back-haul links (e.g., satellite links). There is no spectrum overlapping between the UAV back-haul links and the UAV-user links so that there is no mutual interference. We denote the path loss from UAV $i$ to ground user $u$ as $PL_{iu}$ which follows a commonly adopted model by Al-Hourani \emph{et al.}\cite{al2014modeling}:
\begin{equation}\label{eq.PL}
PL_{iu} = 20\log_{10}{(\frac{4\pi f_cd_{iu}}{c})}+\eta \;\;\; \text{(dB)},
\end{equation}
where $f_c$ denotes the center frequency of the spectrum assigned to user $u$, $d_{iu}$ denotes the 3-D distance between UAV $i$ and user $u$, $c$ denotes the speed of the light, and $\eta$ denotes extra loss taking different values for LoS and non-LoS links. Given Eq. \eqref{eq.PL}, the siginal-to-interference-and-noise ratio (SINR) from UAV $i$ to user $u$ is calculated as
\begin{equation}\label{eq.sinr}
\begin{array}{l}
    SINR_{iu} = \frac{P_tG_{iu}}{n_0+\sum_{j\in {\mathbf{S}^{UAV}_u}\backslash\{i\}}P_tG_{ju}},\\
    \text{where }G_{iu}=10^{-PL_{iu}/20}.
\end{array}
\end{equation}
In Eq. \eqref{eq.sinr}, $P_t$ is the power spectrum density (psd) of UAV transmissions, $n_0$ is the psd of the environment noise, ${\mathbf{S}^{UAV}_u}$ is the set of UAVs that cover user $u$.

Each user requires a minimum throughput of $R_{u}$. Thus, user $u$ will be served by UAV $i$ only when $i\in\mathbf{S}^{UAV}_u$ and the following condition is met:
\begin{equation}\label{eq.access}
W_{iu}\log_2{(1+SINR_{iu})}\ge R_u,
\end{equation}
where $W_{iu}$ is the bandwidth assigned to user $u$ from UAV $i$. According to Eq. \eqref{eq.access}, each user is associated with the UAV which provides the highest SINR with sufficient available bandwidth.  

\subsection{Energy-Related Considerations}\label{subsec.energy}
We consider that each UAV $i$ has an initial battery level $E^i_{0}$. The time horizon is divided into time slots of duration $T$. In time slot $t$, UAV $i$ spends time up to $T_1<T$ to move a distance of $d^i_t\in [0,d_{max}]$ at a constant speed $v$ in the direction of $\alpha^i_t\in[0,2\pi)$, and then hovers in the new position for the remaining time to serve. The power consumption of level flight is given as follows according to \cite{seddon2011basic},  
\begin{equation}\label{eq.level}
P_{level}=\frac{W}{\sqrt{2}\rho A}\frac{1}{\sqrt{v^2+\sqrt{v^4+4V_h^4}}},
\end{equation}
where $V_h=\sqrt{\frac{W}{2\rho A}}$, $W$ is the weight of UAV in Newton ($N$), $\rho$ is the air density, and $A$ is the total area of UAV rotor disks. From Eq. \eqref{eq.level}, it can be inferred that due to speed $v$, the power of level flight is interestingly less than that of hovering. The energy consumption of UAV $i$ in time slot $t$ (denoted as $EC^i_t$) is then represented as
\begin{equation}\label{eq.energy}
EC^i_t = E_{FLT}(v,d^i_t,T) + E_{TX} + E_{OP}(T). 
\end{equation}
According to Eq. \ref{eq.energy}, the energy consumption of a UAV has three components: $i$) energy spent on flying as a function of level speed $v$, flying distance $d^i_t$, and slot duration $T$, $ii$) energy consumed by signal transmission on both UAV-user and UAV back-haul links, and $iii$) energy used for operational cost which is assumed to be proportional to $T$.

Denote the battery residual of UAV $i$ at the end of time slot $t$ as $E^i_t$. When $E^i_t$ is below a threshold $E_{Thre}$, UAV $i$ will quit the network immediately for charging. Denote the altitude of UAV $i$ at the end of time slot $t$ as $H^i_t\in[H_{min},H]$, where $H_{min}$ is the altitude of the UAV charging point. A UAV will stop elevating when it reaches the serving altitude $H$. The variable $H^i_t$ will be used in the case of UAV join-in.

\subsection{Problem Formulation}\label{subsec.obj}
The learning agent aims to find an optimal multi-UAV relocation policy which maximizes the accumulated US scores within a time horizon of $N_T$ time slots, where UAVs may quit or join in the network. The optimization problem is given as follows.
\begin{equation}\label{eq.obj}
    \begin{array}{l}
    \max\limits_{x^i_t,y^i_t}\sum\limits^{N_T}_{t=1}SC_t\\
    s.t. \;\;\;\;0\le x^i_t\le L, \forall i\in\{1,2,\cdots,N_{UAV}\}\\
    \;\;\;\;\;\;\;\;\;0\le y^i_t\le L, \forall i\in\{1,2,\cdots,N_{UAV}\},\\
    \text{where  }\\
        SC_{t}:=\left(\sum_{u\in\mathbf{S}_{ur}}{X^u_t}\right)^\beta.
    \end{array}
\end{equation}
In Eq.\eqref{eq.obj}, $(x^i_t,y^i_t)$ represent the horizontal coordinates of UAV $i$. The US score in time slot $t$ is denoted as $SC_t$ and defined as the function of the total number of users that get served with satisfied throughput requirement. Define $X^u_t\in\{0,1\}$ as an indicator which takes value $1$ when user $u$ is successfully served and $0$ when not. The value of $X^u_t$ is jointly determined by UAV parameters (i.e., the number of serving UAVs, the positions, battery status, altitude), user distribution dynamics, and spectrum access policy. The exponent $\beta>0$ is a factor weighing how much the agent care about the overall user satisfaction based on the number of users successfully served.

With the above formulation, when one UAV is about to be depleted and needs to quit soon, the agent is expected to relocate the serving UAVs ahead of the quit to reduce service holes as much as possible, rather than to react after the UAV quits. When one UAV is joining in the network, the agent is expected to determine its horizontal positions while elevating to the serving height.

\section{Preliminaries}\label{sec.PK}

In the context of a general RL, the agent interacts with the environment by taking an action $A_t$ for the environment state (or observation) $S_t$ at time step $t$. A reward $r_{t+1}$ is then obtained for taking $A_t$ at $S_t$. The learning target is an optimal policy $\pi$ which determines the best action $A$ for every state $S$ that maximizes the expected future return $R$ defined as
\begin{equation}
    R = \sum_{t=0}^{\infty}{\gamma^t r_{t+1}},\;\gamma\in[0,1].
\end{equation}

There are generally two basic categories of RL approaches: value-based and policy-based RL. Q-learning (QL)\cite{han2020} is a basic and representative value-based method. QL achieves the optimal $\pi$ by estimating the value of taking action $A$ at state $S$ which is quantified by the function $Q(S,A)$. The optimal policy $\pi^*$ is obtained as the collection of $A^*=\arg\max\limits_AQ(S,A),\forall S$.
The $Q(S,A)$ function is iterated to a guaranteed convergence according to the following formula,
\begin{equation}
\begin{array}{l}
    Q_{t+1}(S_t,A_t) = \\
    Q_t(S_t,A_t) +\alpha\left[r_{t+1}+\gamma\max\limits_{A}{Q_t(S_{t+1},A)}-Q_t(S_t,A_t)\right],
\end{array}
\end{equation}
where $\alpha$ is the learning rate of the RL agent. Nevertheless, QL has a major drawback that it suffers from the ``curse of dimensionality". A $Q$-matrix needs to be maintained for each state-action pair, which is prohibitive when the state space is extremely large or infinite. This is often the case in communication and networking. To tackle this issue, deep QL (DQL) was developed which exploits a deep neural network (DNN), referred to as deep Q network (DQN), to approximate the $Q(\cdot)$ function\cite{shi2019deep}. The number of inputs of the DQN is equal to the dimension of the state space, and the number of outputs is equal to the cardinality of the action set. Compared to the $Q$-matrix, a DQN reduces the input count to the dimension of the state space and consequently solves the memory anxiety from a large state space.  
The DQN is trained by minimizing the loss function below\cite{hester2018deep}:
\begin{equation}\label{eq.loss}
    \mathcal{L}(\theta_Q)=\mathbb{E}[y_t-Q(S_t,A_t|\theta_Q)]^2,
\end{equation}
where $\theta_Q$ denotes the tunable weights of DQN, $y_t$ is the label value obtained as follows,
\begin{equation}\label{eq.target1}
    y_t = \left\{
    \begin{array}{l}
         r_{t+1},\text{ if }S_t\text{ is a terminal state;}\\
         r_{t+1}+\gamma \max\limits_{A_{t+1}}Q(S_{t+1},A_{t+1}|\theta_Q),\text{ otherwise.}
    \end{array}\right.
\end{equation}

DQL solves the dimension anxiety in state space, but the value-based methods may only apply to problems with low-dimensional discrete action space. The reason is that the value-based methods need to exhaustively search all possible actions to determine the best for a state. Such exhaustive search is difficulty to achieve for a large or infinite action space. When power control or UAV mobility control (as considered in this paper) are involved, the action space is continuous. Discretizing the action space is one possible option, but will lead to prohibitive training complexity and/or non-negligible loss in accuracy. 

\begin{figure}[!ht]
	\centering
	\includegraphics[width=2.5in]{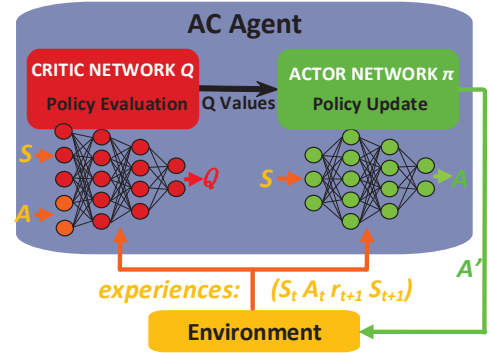}
	\caption{The diagram for AC method.} \label{fig.AC}
\end{figure}

Policy-based methods can well solve the dimension anxiety in the action space. Instead of determining the optimal policy via the $Q(\cdot)$ values, the methods parameterize and optimize the policy $\pi(\theta\mu)$ itself. The $Q(\cdot)$ values may still be used to update the policy parameters $\theta_\mu$, but not for selecting actions directly. Actor-critic (AC) method stands out among all the policy-based methods due to the merit of reducing variance of the policy gradients. A basic AC agent is shown in Fig. \ref{fig.AC}. The agent consists of a critic and an actor, both being DNNs. The critic uses the collected experiences to updates the $Q(\cdot)$ function via updating $\theta_Q$. The actor combines the updated $Q(\cdot)$ values and the experiences to update $\pi$ via updating $\theta_\mu$. The new action $A'$ to be performed is determined by the actor network.

Among all the AC variants, DDPG is one of the best to handle the problem of convergence instability\cite{lillicrap2015continuous}. Specifically, DDPG exploits target networks for both the critic network ($Q(S,A|\theta_Q)$) and actor network ($\mu(S|\theta_\mu)$). The target networks, denoted as $Q'(S,A|\theta_{Q'})$ and $\mu'(S|\theta_{\mu'})$, have the same setup and initialization as $Q(S,A|\theta_Q)$ and $\mu(S|\theta_\mu)$, respectively, but are updated much slowly in each time step:
\begin{equation}\label{eq.target}
    \theta_{Q'} = \tau\theta_Q+(1-\tau)\theta_{Q'},\;\;\;\;\;\;
    \theta_{\mu'} = \tau\theta_\mu+(1-\tau)\theta_{\mu'}
\end{equation}
where $\tau<<1$. The target networks are used to update the label value $y_t$ in Eq. \eqref{eq.loss}. Correspondingly, Eq. \eqref{eq.target1} is re-written as
\begin{equation}\label{eq.losstarget}
    y_t = \left\{
    \begin{array}{l}
         r^{t+1},\text{ if }S_t\text{ is terminal state;}\\
         r_{t+1}+\gamma Q'(S_{t+1},\mu'(S_{t+1}|\theta_{\mu'})|\theta_{Q'}),\text{ otherwise.}
    \end{array}\right.
\end{equation}
The slow update of the target networks prevents a bad $y_t$ from being generated due to a bad deviation in $\theta_{Q}$ or $\theta_{\mu}$, thus significantly stabilizing the convergence.

Using the updated $Q$ values, the actor network is updated as follows:
\begin{equation}\label{eq.actor}
    \begin{array}{l}
        \nabla_{\theta_\mu}J \approx \mathbb{E}[G_{a}G_{\mu}],\\
        \text{where }G_{a}=\nabla_{\mu(S|\theta_\mu)}Q(S,A|\theta_Q),\\
        \;\;\;\;\;\;\;\;\;\;G_{\mu}=\nabla_{\theta_\mu}\mu(S|\theta_\mu).
    \end{array}
\end{equation}
\section{Learning Algorithm Design for Proactive Self-Regulation Strategy}\label{sec.Algorithm}
The design of the proposed PSR-APC approach is detailed in this section. The DDPG agent is implemented in a centralized server, which communicates regularly with all the UAVs via their backhaul links. 
During training, the agent keeps collecting the interaction experiences between UAVs and the network environment and updating both critic and actor networks. When the training is complete and the strategy is executed, the well-trained actor network sequentially collects the UAV information (states) as inputs and outputs flying instructions (actions) to each UAV in each time step. These movements collectively result in an optimal set of UAV trajectories to maximize the accumulative US score within the considered time horizon. 

In addition, we consider that after one UAV quits or joins in, all the serving UAVs have sufficient time to reach the new optimal positions before another lineup change happens. Therefore, the case of multiple lineup changes can be regarded as multiple cases of single lineup change. In such a case, a different trained agent for each lineup change will be exploited sequentially to regulate the network in the considered time horizon.

In the following, we elaborate the design from the aspects of states, actions, reward function, state transitions, tune-ups, and parallel computing. The cases of UAV quit and join-in are both considered.

\subsection{State Space}
\subsubsection{\textbf{Case of UAV Quit}}
The timing of UAV quit and the resultant UAV movements are closely dependent on the battery level of the UAVs. Hence, the learning states will include UAV positions and battery residual of each UAV.

The UAV positions directly determine the number of users that get successfully served in each time step. As the UAVs fly at a fixed height when serving, only the 2-D coordinates $(x^i_t,y^i_t)$, $\forall i\in\mathbf{S}_{UAV}$ need to be considered at time step $t\le N_T$. The movements of UAVs are limited within the target area $\mathbf{A}$, i.e., $x^i_t,\;y^i_t\in[0,L]$.

The battery residual of UAVs $\{E^i_t\}$ is a key conditional factor. It has little impact on the UAV movements when the battery level of all the UAVs is adequate. Yet when any $E^i_t$ falls close to $E_{Thre}$ (i.e., any UAV is running out of battery and about to quit), this factor should have significant impact on the UAV movements. The best timing of enabling the significance of $\{E^i_t\}$ will be learnt by the DDPG agent. Moreover, $E^i_t$ is bounded within $[E_{Thre},E^i_0]$.

Collectively, the formal state vector of the designed learning approach is defined as $S_t=[x^1_t,\cdots,x^{N_{UAV}}_t,y^1_t,\cdots,$ $y^{N_{UAV}}_t,E^1_t,\cdots,E^{N_{UAV}}_t]$, with cardinality of $3N_{UAV}$.
\subsubsection{\textbf{Case of UAV Join-In}} A UAV is considered to start serving only when it reaches the serving altitude $H$. Similar to the battery residual of UAVs in the case of UAV quit, UAV altitude is the key factor in this case which determines the timing of proactive UAV relocation. The existing UAVs will bide their time until the joining UAV is about to reach the serving altitude. While elevating to the serving altitude, the joining UAV needs to adjust its horizontal position since where to join the UAV network is critical to maximizing the accumulative US score. 

Hence, the formal state vector of the designed learning approach is defined as $S_t=[x^1_t,\cdots,x^{N_{UAV}}_t,y^1_t,\cdots,$ $y^{N_{UAV}}_t,H^1_t,\cdots,H^{N_{UAV}}_t]$, with cardinality of $3N_{UAV}$. Based on the collected experiences, the agent will learn the best period for $\{H^i_t\}$ to take effect.

\subsection{Action Definition}\label{subsec.action}
The action set of the APC-PSR approach is the same for both cases. As a centralized agent controls the movements of all the UAVs, the collective actions from all the UAVs form the agent action $A_t$ in time step $t$. The action $A^i_t$ of UAV $i$ has two dimensions: moving direction $\alpha^i_t\in[0,2\pi)$ and moving distance $d^i_t\in[0,d_{max}]$. In each time step, one UAV could either keep hovering still or move in any direction for a maximum distance $d_{max}$. Thus the formal action vector of the proposed APC-PSR approach is defined as $A_t=[\alpha^1_t,\cdots,\alpha^{N_{UAV}}_t,d^1_t,\cdots,d^{N_{UAV}}_t]$, with cardinality of $2N_{UAV}$.

\subsection{Reward Function Design}
Both the cases of UAV quit and join-in share the same reward function design. Let $r_t$ denote the reward at time step $t$. To align with the maximization objective in Eq. \eqref{eq.obj}, $r_t$ is designed as a function of the instantaneous US score in step $t$, i.e., $SC_t$:
\begin{equation}\label{eq.reward}
    r_t = \left(\frac{\sum_{u\in\mathbf{S}_{ur}}{X^u_t}}{N_u}\right)^\beta=\frac{SC_{t}}{(N_u)^\beta}.
\end{equation}
In \eqref{eq.reward}, the instantaneous US score $SC_t$ is divided by $(N_u)^\beta$. Empirically speaking, keeping the absolute value of the instantaneous reward within $1$ may result in better convergence. In addition, when $\beta>1$, the reward difference between different ($\sum_{u\in\mathbf{S}_{ur}}{X^u_t}$) values is amplified. This promotes the agent to act in advance when the UAV lineup is about to change. However, $\beta$ cannot be too large as $\beta\ge3$ has been shown ending up with lower converged values in our preliminary simulations. Moreover, under this design, maximizing the accumulated reward is equivalent to maximizing the accumulated US scores within $N_T$ time steps. 

An alternative design of reward function is to give negative rewards as a punishment when any UAV move out of the boundaries\cite{liu2018energy}. The reward function will be something like:
\begin{equation}
    r_t = \left\{
    \begin{array}{l}
    \left(\sum_{u\in\mathbf{S}_{ur}}{X^u_t}/{N_u}\right)^\beta,\;\;\text{if inside boundaries}\\
    P,\;\;\text{otherwise}
    \end{array}
    \right.
\end{equation}
where $P$ can be a negative constant or variable proportional to the number of UAVs crossing the boundaries. During training, when one UAV moves out of boundaries, its current movement will be cancelled. A negative reward will be issued for taking the current action $A_t$ at the current state $S_t$. With such a design, all the episodes will have a fixed number of $N_T$ time steps. Reasonable as this design is, it may make convergence more difficult. This is because in a good reward design with both positive and negative rewards, the negative rewards need to ``combat" the positive ones closely during the training for better convergence performance and speed. However, the relative ratio between the positive and negative rewards keeps changing during the training, making it more challenging and computationally complex to achieve satisfying convergence.

\subsection{State Transition Definition}

In either case, a state is a terminal state if at least one of the two conditions is met: $i$) when any UAV moves outside the boundaries of the target area, i.e., $x^i_t<0$, $y^i_t<0$, $x^i_t>L$, or $y^i_t>L$; or $ii$) when $N_T$ time steps are completed. The current episode will end When the terminal state is reached and a new one will start.
    
Due to the dynamic UAV lineup change, the number of UAVs in the network may change accordingly. This results in that the dimension of the \emph{actual} state-action space to explore during the training will vary after one UAV quits or joins the network. 
\subsubsection{\textbf{Case of UAV Quit}}Consider that UAV $i$ quits the network at time step $t_q$. Then $x^i_t$, $y^i_t$, and $E^i_t$ will stay unaltered for any $t>t_q$. Whatever actions $\alpha^i_t$ and $d^i_t$ ($t>t_q$) are selected, the positions and battery residual of UAV $i$ will never be updated. In other words, the actual dimension of the explorable state space is reduced from $3N_{UAV}$ to $3(N_{UAV}-1)$. At the same time, UAV $i$ will not be considered in the reward calculation after $t_q$.
\subsubsection{\textbf{Case of UAV Join-in}}Suppose UAV $i$ completes charging and is ready to take place to join the network at time step $t_c$. When $t<t_c$, $x^i_t$, $y^i_t$, and $H^i_t$ will stay unaltered. Suppose UAV $i$ elevates to the serving altitude at time step $t_s$. When $t_c\le t<t_s$, UAV $i$ is excluded from the instantaneous reward calculation, but its horizontal positions ($x^i_t$, $y^i_t$) will change towards the optimal position to maximize the instantaneous reward when formally joining the network. A constant elevation distance $h$ per time step will be used. 

\subsection{Training Tune-Ups}
\subsubsection{\textbf{Tune-Ups for Neural Network Training}}Both the critic and actor networks are DNNs. We design both networks to be just complex enough to accurately learn the nonlinear mappings between inputs and outputs while preventing overfitting. Both DNNs contain 2 fully connected hidden layers with 400 and 300 hidden nodes, respectively. To bound the actions as designed in Subsection \ref{subsec.action}, we employ \emph{tanh} and \emph{scaling} layers in the actor network. In both networks, ReLU function is used as the activation function, and $L_2$ regularization is adopted to suppress over-fitting. The learning rates for updating both $\theta_Q$ and $\theta_\mu$ is $10^{-4}$. Although a larger learning rate may expedite convergence, it more likely leads to convergence instability or sub-optimum. We choose the mini-batch size for DNN training to be $512$ which is comprises between computational efforts and variance reduction of the gradients of the loss functions. Input normalization is also enabled for faster convergence.

\subsubsection{\textbf{Tune-Ups for RL Training}}During RL training, both target networks $Q'(S,A|\theta_{Q'})$ and $\mu'(S|\theta_{\mu'})$ are updated slowly at a rate $\tau=0.001$. The discount factor $\gamma$ is set to 0.9. A higher $\gamma$ will force the agent to account more of the future rewards, thus making the converging more difficult. In addition, DDPG adopts an exploration algorithm where the output of the actor network is add with a random noise of zero mean and decaying variance over time steps. In our implementation, the initial variance is 0.6 and decays at a rate of 0.9995. Experience replay is used with sufficient buffer to contain all the experiences. Insufficient buffer may make the agent lose valuable experiences at early stage if one does not know well which experiences to drop, which will cause notable convergence instability or even divergence. 

\subsection{Parallel Computing}
One major challenge during the training is how to fully explore the state-action space to promote action taking ahead of the lineup change. Failing to do so will lead to convergence to a sub-optimal (sometimes even bad) result, which is often the case in our early simulations. The reason of insufficient exploration is mainly two-folds. First, although experience replay randomly sample experiences from the entire buffer to train the DNNs, the sampled experiences in one mini-batch will inevitably have some correlation due to the Markov nature of RL. Correlation among training examples of DNN will harm the learning accuracy. Second, the dimension of the explorable state space changes accordingly when the UAV lineup changes. Such a change during training often leads to no UAV relocation after the change or no proactive movement ahead of the change. 

\begin{figure}[!ht]
	\centering
	\includegraphics[width=2.8in]{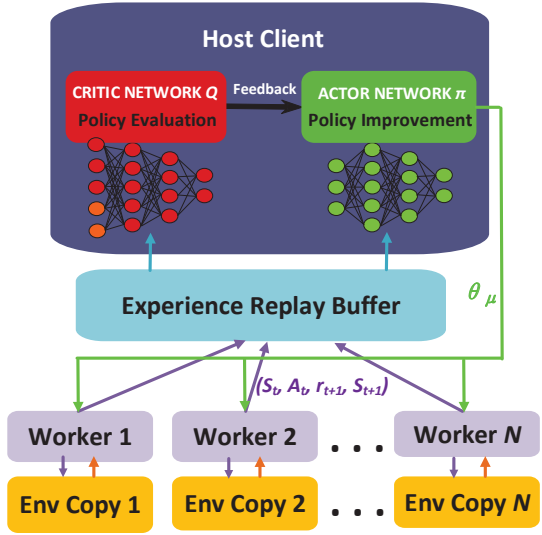}
	\caption{The diagram for asynchronous parallel computing (APC) of DDPG algorithm.} \label{fig.APC}
\end{figure}

Increasing the random noise added to the output of the actor network does not help in our case. Inspired by the asynchronous advantage actor critic (A3C) \cite{mnih2016asynchronous} algorithm, we propose to use the structure of asynchronous parallel computing (APC) to promote exploration, as shown in Fig. \ref{fig.APC}. 
The structure contains a host client and multiple parallel workers. The host client maintains a unified pair of critic and actor networks and periodically updates both networks from the collected experiences. The parallel workers are mutually independent, each interacting with an independent copy of the same environment. In A3C, each worker has its own set of network parameters and send the \emph{gradients} of policy loss to the host client. But in APC, the parallel workers share the same set of policy parameters from the host client and upload their own \emph{experiences} to the host client to update the unified neural networks. In our implementation, each worker uploads the experiences upon completion of the current episode and immediately receives updated policy parameters from the host, thus being asynchronous from each other.

\begin{algorithm}[!ht]
\caption{PSR-APC Approach: Host Client Side (UAV Quit/Join-in)}
\label{Alg:algorithm1}
\begin{algorithmic}[1]
\State{/*\textbf{Host Client}*/}
\State{Randomly initialize critic network $Q(S,A|\theta_Q)$ and actor network $\mu(S|\theta_\mu)$;}
\State{Initialize the target networks $Q'(S,A|\theta_{Q'})$ and $\mu'(S|\theta_{\mu'})$ with the same weights: $\theta_{Q'} := \theta_{Q}, \theta_{\mu'}:=\theta_{\mu}$;}
\While{Not all workers complete all episodes}
    \If{Receive experience set $\mathbf{Ep}$ from worker $k$}
        \State{Store $\mathbf{Ep}$ into experience replay buffer $B$;}
        \State{Send $\theta_\mu$ to worker $k$;}
    \EndIf
    \State{Sample a mini-batch of experiences from $B$;}
    \State{Update $\theta_Q$ according to Eq. \eqref{eq.loss}\eqref{eq.losstarget};}
    \State{Update $\theta_{\mu}$ according to Eq. \eqref{eq.actor};}
    \State{Update $\theta_{Q'}$ and $\theta_{\mu'}$ according to \eqref{eq.target};}
\EndWhile
\end{algorithmic}
\end{algorithm}
\begin{algorithm}[!ht]
\caption{PSR-APC Approach: Parallel Worker Side (UAV Quit/Join-in)}
\label{Alg:algorithm2}
\begin{algorithmic}[1]
\State{/*\textbf{Parallel Worker}*/}
\For{episode := $1,\cdots,N$}
    \State{Obtain the initial state $S_1$, IsTerminal := False;}
    \For{epoch $t$ := $1,\cdots,N_T$}
        \State{$A_t=\mu(S_t|\theta_\mu)+\mathcal{N}$, where $\mathcal{N}$ is stochastic noise $\text{ }\text{ }\text{ }\text{ }\text{ }\text{ }\text{ }\text{ }$ with zero mean and decaying variance over $t$;}
        \State{Execute $A_t$ and observe next state $S_{t+1}$;}
        \For{UAV $i$ := $1,\cdots,N_{UAV}$}
            \State{/*Case of \emph{UAV Quit}*/}
            \If{$E^i_t<=E_{Thre}$}
                \State{$S^i_{t+1}:=S^i_t$, where $S^i_t$=$\{x^i_t,y^i_t,E^i_t\}$;}
                \State{Exclude UAV $i$ when calculating $r_{t+1};$}
            \Else
                \State{Obtain $S^i_{t+1}$ according to $S^i_t$ and $A_t$;}
            \EndIf
            \State{/*Case of \emph{UAV Join-In}*/}
            \If{$H^i_t<H$}
                \State{Exclude UAV $i$ when calculating $r_{t+1}$;}
                \State{$H^i_{t+1}=\min\{H^i_t+h,H\}$;}
            \EndIf
            \State{Obtain $\{x^i_{t+1},y^{i_{t+1}}\}$ according to $S^i_t$ and $A_t$;}
            \State{/*Shared codes begin*/}
            \If{UAV $i$ goes out of boundaries}
                \State{Cancel the movement of UAV $i$;}
                \State{IsTerminal := True;}
            \EndIf
        \EndFor
        \State{Calculate $r_{t+1}$;}
        \If{IsTerminal==True}
            \State{Break;}
        \EndIf
    \EndFor
    \State{Send experiences in this episode to host client;}
    \State{Obtain updated $\theta_\mu$ from host client;}
\EndFor
\end{algorithmic}
\end{algorithm}

Since independent workers interact with different copies of the same environment, the experiences from each worker will be independent. In this way, the correlation among the sampled experiences in a mini-batch are significantly diluted, thus leading to potentially better network training performance. Note that APC itself does not increase the convergence speed in our case as the computational complexity of neural network training is much higher than that of simulating the environment. As a summary, Algorithms \ref{Alg:algorithm1} and \ref{Alg:algorithm2} display the pseudo-codes of the proposed APC-PSR approach for both cases of UAV quit and join-in.

\section{Proactive Self-Regulation with Dynamic User distribution}\label{sec.Dynamic}

The above section considers a \emph{fixed} user distribution in the entire time horizon. But it may be more practical if \emph{dynamic} user distribution is considered. With dynamic user distribution in the considered time horizon, the optimal UAV positions may have to change from time to time in order to maximize the accumulated US score. This is more challenging compared to the case of the fixed distribution. In such a situation, the optimal self-regulation of UAVs is more towards the optimal UAV trajectory design, with additional proactive response to the UAV lineup change.


In this problem, how to determine the optimal trajectories for all the UAVs according to the dynamically changing user distributions is the key. Some existing works\cite{chen2017caching,liu2019trajectory} have considered time-varying user distributions under the RL framework. In these works, time-varying user mobility patterns are first predicted using either echo state networks (ESNs) or Long-Short Term Memory (LSTM), based on which (multi-agent) QL is employed to achieve the optimal UAV trajectories. Nevertheless, the trajectories are obtained in such a way that optimal positions in each time slot are first derived by RL algorithms given a slot-specific user distribution, and then stringed together into the trajectories. These methods have to train the RL agent(s) once every time slot, instead of training once for the entire time horizon with dynamic user distribution. This may incur high training complexity if a large number of time slots are involved. Different from the above works, we incorporate the time slot $t$ as one dimension of the state space, so that the obtained policy is determined not by one specific distribution in a particular time step, but by a dynamic distribution along the entire time horizon. In this manner, the agent only needs to be trained once over the entire time horizon, and ends up with a time-aware optimal policy that may take different actions at different time slots even if the rest of the states are the same.

\begin{figure*}[t]
\centering
\begin{tabular}{llll}
\subfigure[Snapshot 1] {\includegraphics[height=1.3in,width=1.45in,angle=0]{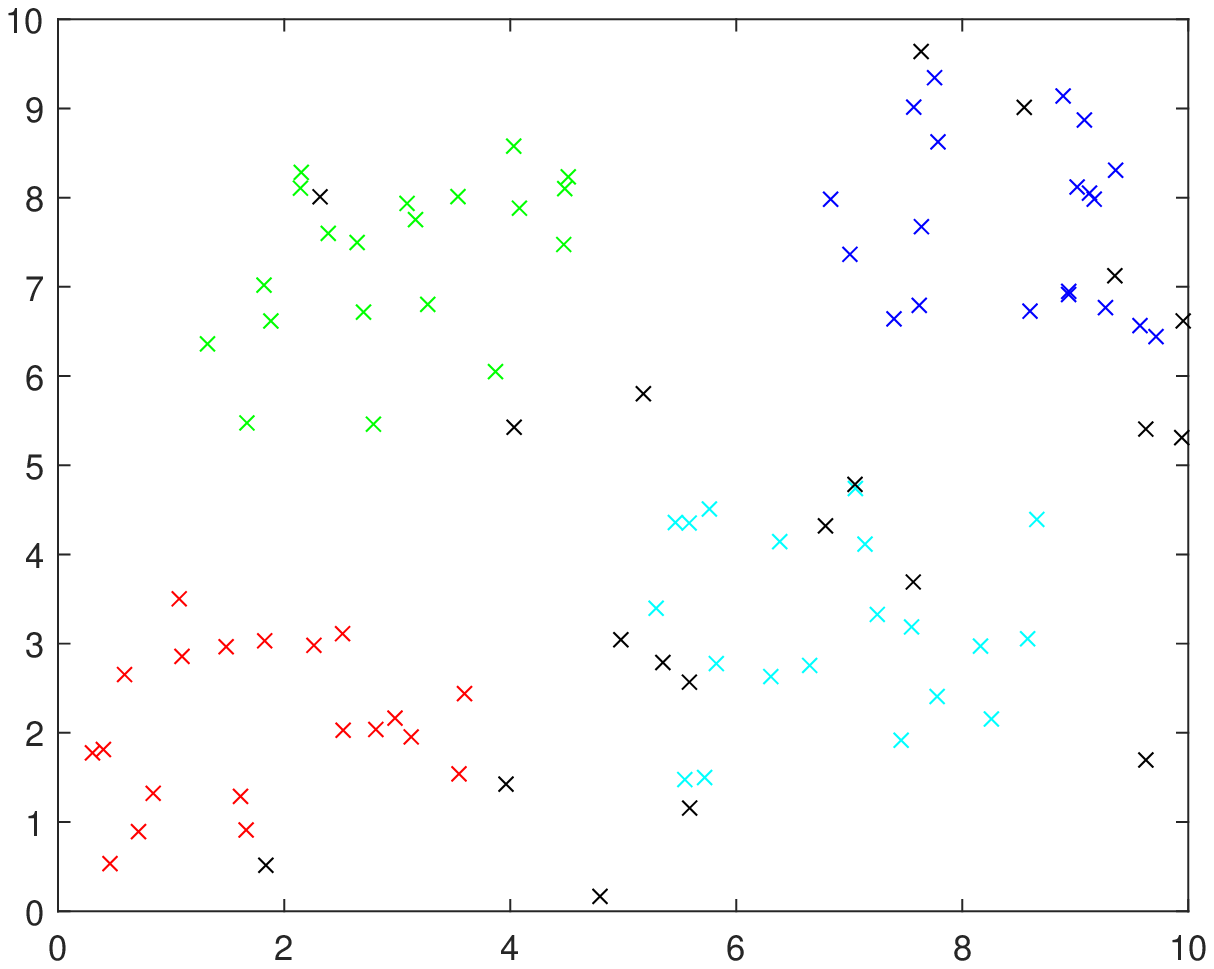}\label{Fig.1}} &
\subfigure[Snapshot 2] {\includegraphics[height=1.3in,width=1.45in,angle=0]{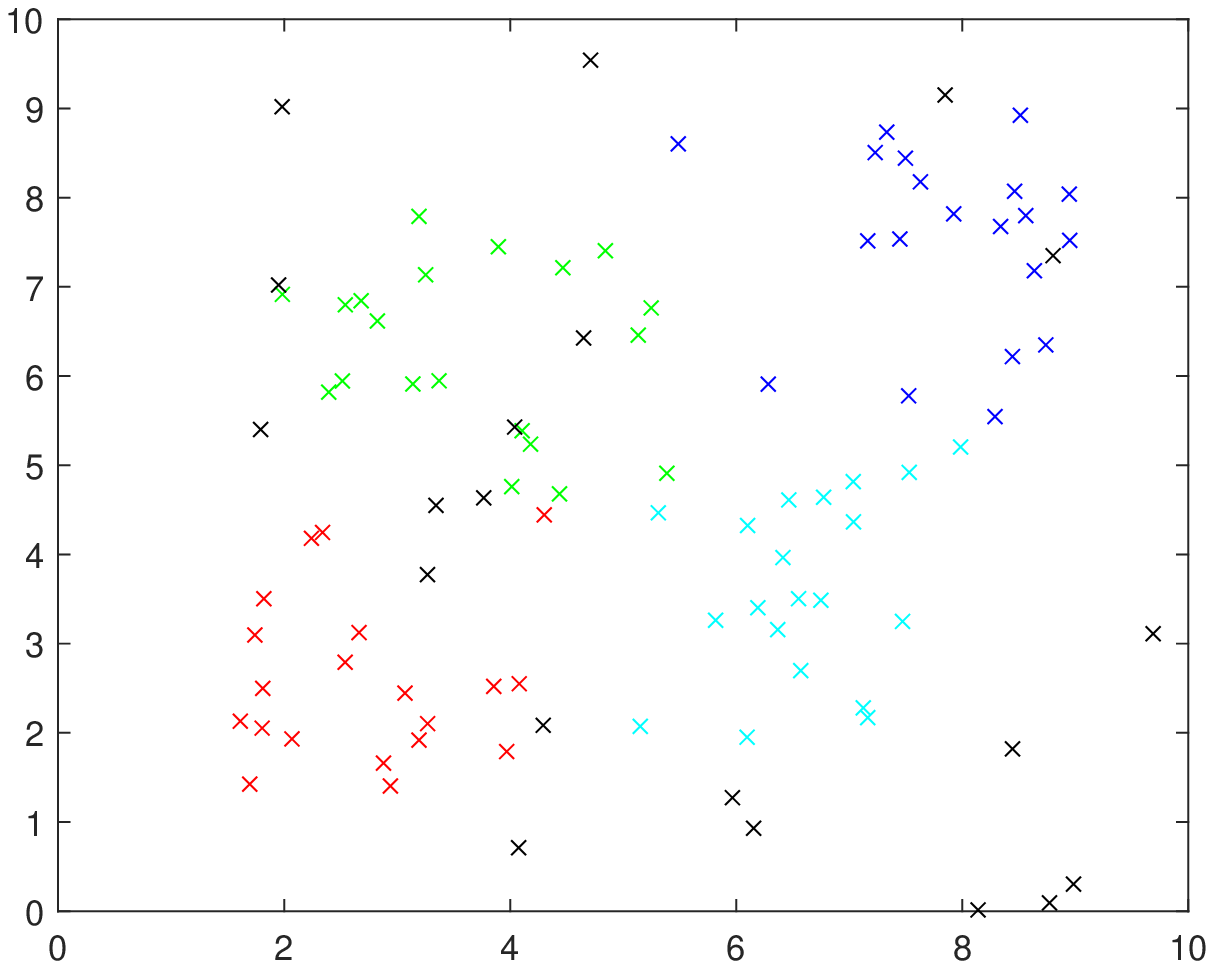}\label{Fig.2}} &
\subfigure[Snapshot 3] {\includegraphics[height=1.3in,width=1.45in,angle=0]{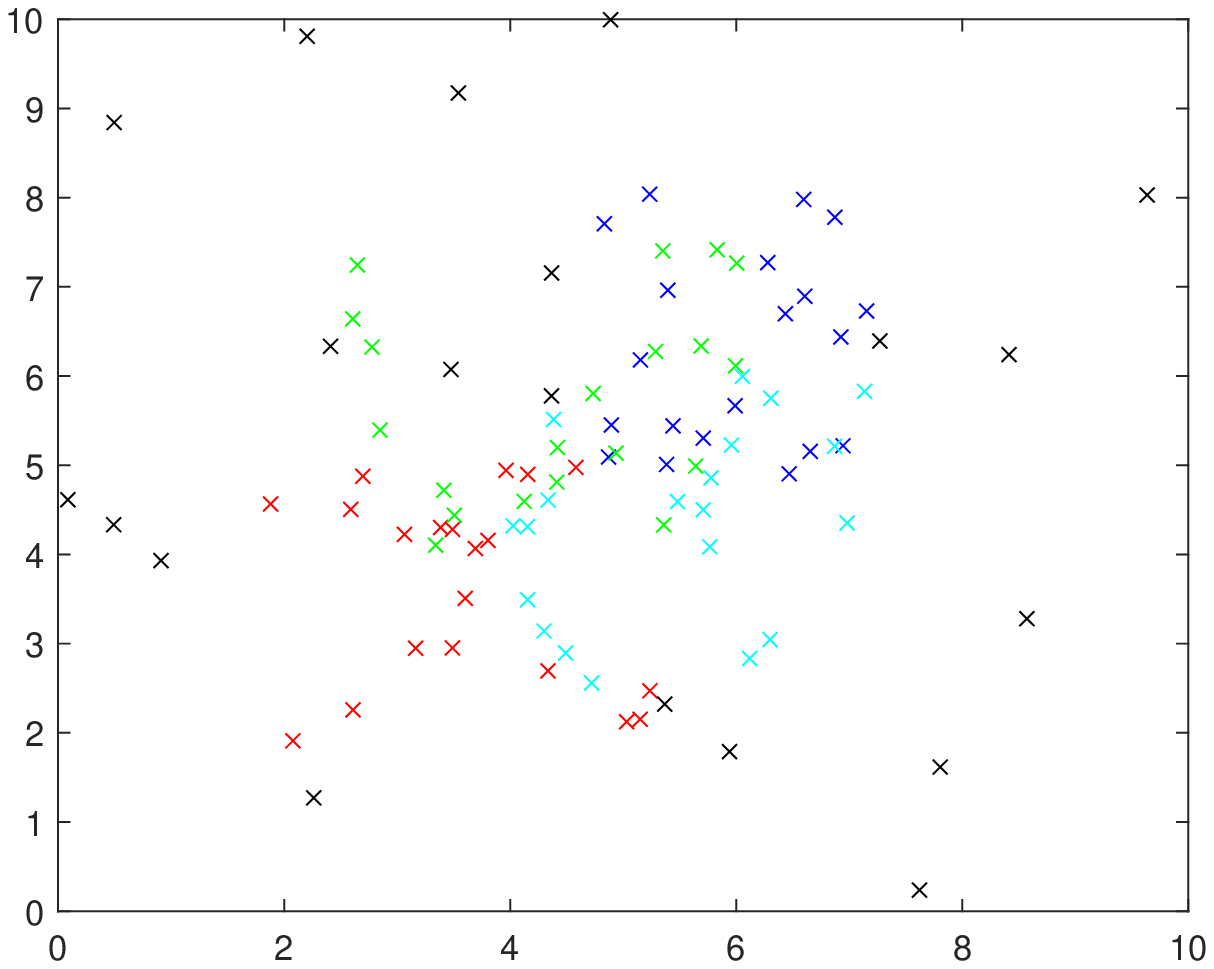}\label{Fig.3}} &
\subfigure[Snapshot 4] {\includegraphics[height=1.3in,width=1.45in,angle=0]{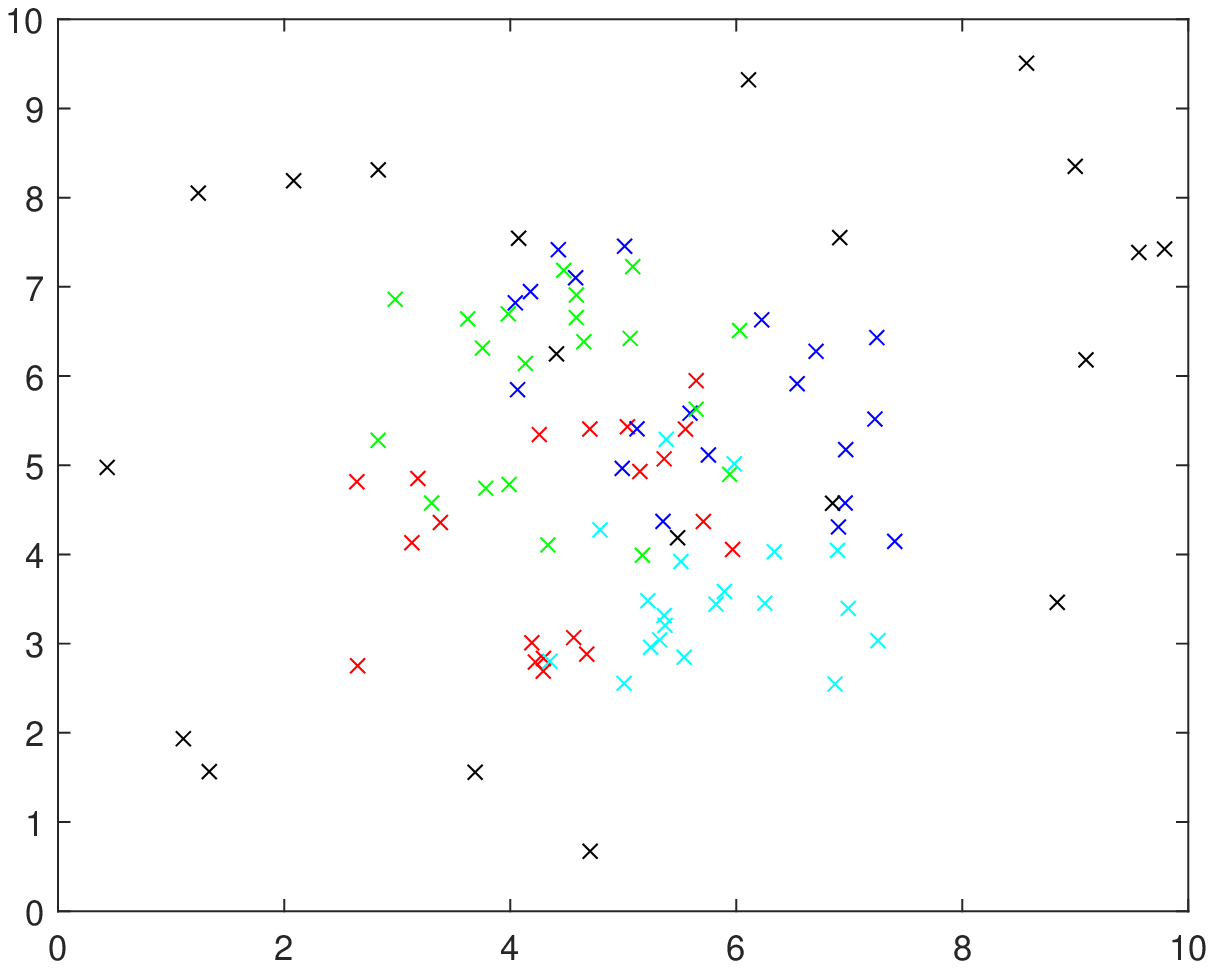}\label{Fig.4}}
\end{tabular}
\caption{\small{Illustration of time-varying user distribution: Snapshots of different time slots. Users are represented by dots of different colors. Users of black are uniformly distributed throughout $\mathbf{A}$, and users of each other color belongs to one hot spot. From snapshots 1 to 4, hot spots move from being scattered to being overlapped.}}\label{fig.distribution}
\end{figure*}

A simplified model on dynamic user distribution is exploited to investigate the learning performance of the APC-PSR approach on the time-variability of the user distribution. Instead of specifying mobility models for individual users, a different trace is considered for each hot spot center. That is, the center of each hot spot follows a different race to move in the target area $\mathbf{A}$ with time $t$. The percentage of users in proximity to each hot spot can be either fixed or moderately varying. Fig. \ref{fig.distribution} illustrates an example of moving hot spots and the corresponding user distribution. It can be seen that there are 4 hot spots initially located at 4 corners of $\mathbf{A}$. During $N_T$ time steps, the 4 hot spots first move towards the center of $\mathbf{A}$, i.e.,  ($L/2$,$L/2$), stop when reaching a certain distance from the center, stay for a period, and finally move back to where they were initially. Such a disperse-gather-disperse procedure can be used to simulate some realistic scenarios such as when users commute between residencies and a central business district during workdays\footnote{In such a scenario, the duration of one time step needs to be scaled up to the order of minutes.}. 

As the user distribution is changing, the agent may take different actions even if the UAVs are in the same positions and energy/altitude status at different time steps. Therefore, the design of APC-PSR approach needs to be modified to include time as one of the states. Hence we re-define the states of the proposed APC-PSR approach as: $S_t=[x^1_t,\cdots,x^{N_{UAV}}_t,y^1_t,\cdots,y^{N_{UAV}}_t,E^1_t,$ $\cdots,E^{N_{UAV}}_t, t]$ (case of UAV quit), or $S_t=[x^1_t,\cdots,x^{N_{UAV}}_t, y^1_t,\cdots,y^{N_{UAV}}_t,$ $H^1_t,\cdots,H^{N_{UAV}}_t,t]$ (case of UAV join-in), both with cardinality of $3N_{UAV}+1$. 

Note that despite the example user distribution, the proposed approach is deemed to be applicable to any kind of distribution dynamics as long as the distribution can be considered invariant within one time step.

\section{Numerical Results}\label{sec.Simulation}
Numerical results are presented in this section to demonstrate the performance of the proposed APC-PSR approach. 

\subsection{Simulation Setup}
The target area is a $10\times10$ unit square with each unit being 100 meters. The simulations are conducted using Reinforcement Learning Toolbox of Matlab 2020a on a Windows 10 server with Intel Core i7-7700 CPU @ 3.60GHz and 16GB RAM. The training has maximum 10000 episodes, each having up to 100 time steps. The trained agents are tested for a period of $N_T=100$ time steps. In addition, we consider the transmission-related power of UAVs negligible compared to the propulsion power\cite{zeng2017energy}. Table \ref{Table:notation} summarizes the main parameters below. Note that $unit\cdot s$ in the table indicates that the value is a product of power (1 power unit = 9.428W according to Eq. \eqref{eq.level}) and time (unit is second).

\begin{table}[!ht]
\footnotesize
\centering
\renewcommand{\arraystretch}{1}

\begin{tabular}{!{\vrule width0.8pt}l|l!{\vrule width0.8pt}}\Xhline{0.8bp}
\multicolumn{1}{!{\vrule width0.8pt}c|}{\gape{\bfseries Parameters}} & \multicolumn{1}{c!{\vrule width0.8pt}}{\gape{\bfseries Values}} \\ 



  \hline
  \rowcolor[gray]{0.9}
  Default number of users $N_u$ & 100\\
  Default number of UAVs $N_{UAV}$ & 5\\
  \rowcolor[gray]{0.9}
  UAV level speed $v$ & 40$km/h$\\
  \rowcolor[gray]{0.9}
  UAV max. elevation speed & 14.4$km/h$\\
  UAV weight $W$, air density $\rho$& 4$kg\times$9.8$m/s^2$, 1.225$kg/m^3$\\
  \rowcolor[gray]{0.9}
  Total area of rotor disks $A$ & 0.18$m^2$\\
  UAV height $H$, apenture angle $\theta$  & 3 units, $60^o$ \\
  \rowcolor[gray]{0.9}
  Max. distance per epoch $d_{max}$ & 1 unit \\
  Spectrum center frequency $f_c$  & 2GHz\\
  \rowcolor[gray]{0.9}
  Spectrum access technology & LTE with resource blocks (RBs)\\
  Spectrum and RB bandwidth & 4.5MHz and 180kHz\\
\rowcolor[gray]{0.9}
  psd of transmission and noise& -49.5dBm, -174dBm \\
  Required user throughput $R_u$ & 250kbps\\
  \rowcolor[gray]{0.9}
  LOS path loss parameter $\eta$&1dB \\
  Time duration per epoch $T$&10s\\
  \rowcolor[gray]{0.9}
  Max. UAV moving (communica- & 9s (1s)\\
  \rowcolor[gray]{0.9}
  tion) time per epoch $T_1$ ($T-T_1$)&~~ \\
  Factor of US score $\beta$&2 \\
  \rowcolor[gray]{0.9}
  Energy threshold to quit $E_{Thre}$ & 150 unit$\cdot$s\\

  \hline
  \end{tabular}
\caption{Summary of Main Parameters}\label{Table:notation}
\end{table}

The learning converges to a narrow range instead of a fixed value in most of the simulated cases. For the sake of better presentation, we smooth the convergence curve of the episode reward by averaging over the latest 100 episodes. Intermediate agents with good episode rewards are saved during training and compared during tests to determine the best trained agent.

\subsection{Simulation Results}\label{subsec.simuresults}
\subsubsection{\textbf{Case without UAV or user dynamics}}
\begin{figure}[!ht]
	\centering
	\includegraphics[width=3.0in]{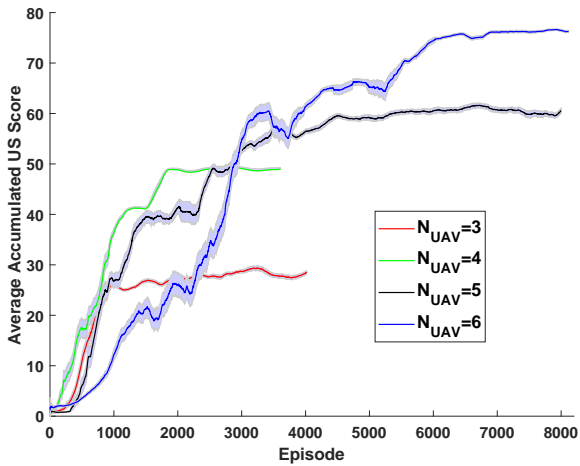}
	\caption{Convergence with 95$\%$ credit interval for different $N_{UAV}$ without UAV or user dynamics as a benchmark} \label{fig.Converge}
\end{figure}

We first simulate the cases without any UAV lineup or user distribution change to get a reference of optimal UAV positions under different $N_{UAV}$. Fig. \ref{fig.Converge} shows the convergence performance of the accumulated US scores. The initial positions of all the UAVs are the evenly separated dots on a circle centered at (5,5). It can be seen that for all the simulated $N_{UAV}$ values, the accumulated US scores eventually converge, with larger $N_{UAV}$ taking more training episodes. The reason is straightforward: the more UAVs there are, the larger dimension of the state-action space has, thus requiring more time to fully explore and exploit. In addition, the converged accumulated US score is smaller for smaller $N_{UAV}$, which aligns with the fact that more UAVs can cover more users until the target area is saturated with UAVs. However, as $N_{UAV}$ increases, the increment in the number of served users reduces according to Table \ref{Table:2}. 
\begin{table}[!htbp]
\centering
\begin{tabular}{c|c|c|c|c}
\hline
$N_{UAV}$ & 3 & 4 & 5 & 6\\
\hline
Number of served users& 56 & 71 & 80 & 88\\
\hline
Increment & - & 15 & 9 & 8\\
\hline
\end{tabular}\caption{Maximum number of served users with $N_{UAV}$.}\label{Table:2}
\end{table}

\begin{figure}[!ht]
	\centering
	\includegraphics[width=3.0in]{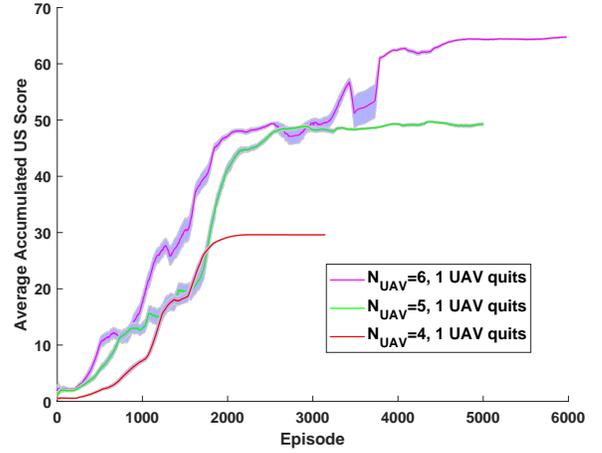}
	\caption{Convergence performance with 95$\%$ credit interval for the case of UAV quit.} \label{fig.UAVQuitConverge}
\end{figure}
\subsubsection{\textbf{Case of UAV quit}}
\begin{figure}[!ht]
\begin{tabular}{l}
\subfigure[$N_{UAV}=6$] {\includegraphics[width=2.8in]{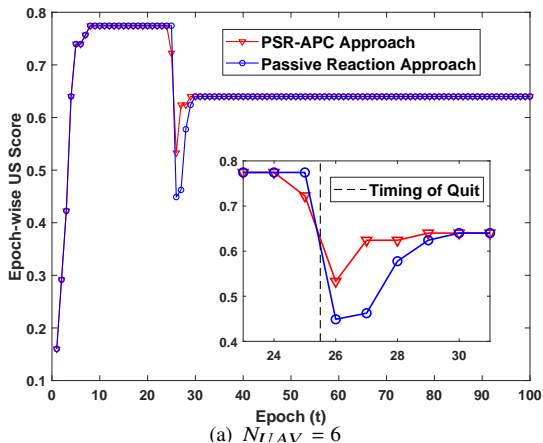}\label{Fig.N6}}\\ 
\subfigure[$N_{UAV}=5$] {\includegraphics[width=2.8in]{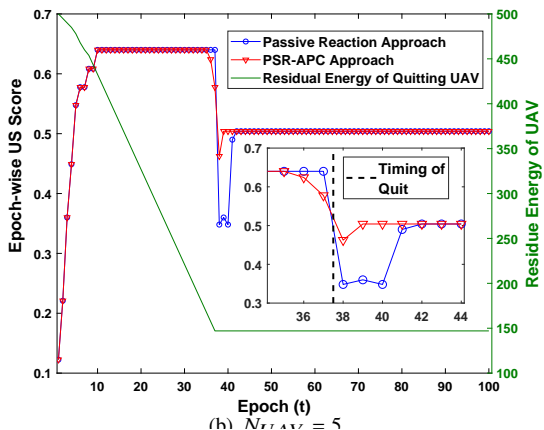}\label{Fig.N5}} \\
\subfigure[$N_{UAV}=4$] {\includegraphics[width=2.8in]{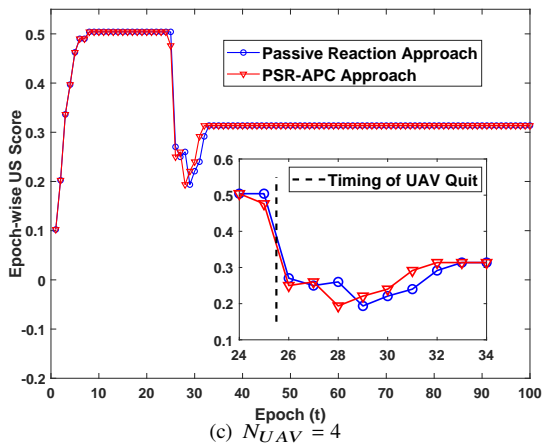}\label{Fig.N4}}
\end{tabular}
\caption{\small{Case of UAV quit: Epoch-wise reward comparison between the PSR-APC approach and the passive reaction approach.}}\label{fig.Epoch_Quit}
\end{figure}

We then simulate the case of UAV quit where $N_{UAV}$ UAVs start off in the beginning and 1 UAV quits within the considered period. The UAVs are initially positioned at the optimal locations obtained via Fig. \ref{fig.Converge}. Although multiple UAVs may quit during the period, we consider that when one UAV quits, the remaining UAVs will reach the new optimal positions before another UAV quits. Hence the case of multi-UAV quit can be broken into multiple cases of single-UAV quit. The convergence performance is presented in Fig. \ref{fig.UAVQuitConverge}. It can be seen that it takes more episodes for larger $N_{UAV}$ to converge. Then, the optimal epoch-wise US scores are shown in Fig. \ref{fig.Epoch_Quit}. As a comparison to the proposed PSR-APC approach, a passive reaction approach is also simulated, which only relocates the remaining UAVs passively after one UAV quits the network.


\begin{figure}[!ht]
	\centering
	\includegraphics[width=3.3in]{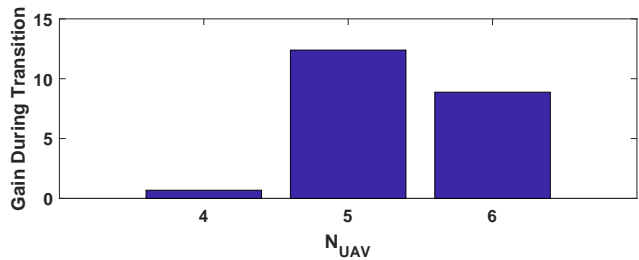}
	\caption{\small{Case of UAV quit: Gain ($\%$) of PSR-APC over the passive reaction approach in accumulated US scores during transition to the new optimal UAV positions.}} \label{fig.Gain_Quit}
\end{figure}

The epoch-wise US scores in Fig. \ref{fig.Epoch_Quit} are obtained by combining the agents achieved in Fig. \ref{fig.Converge} and Fig. \ref{fig.UAVQuitConverge}. The UAVs start from the circular positions, then to the optimal positions maximizing the epoch US score; a UAV then quits the network and the remaining UAVs are finally relocated to the new optimal positions. It can be observed that for all the simulated $N_{UAV}$, the epoch US score first increases to a maximum as the UAVs are heading to the optimal positions. When a UAV quits the network, the epoch US scores drop dramatically due to the service holes caused by the quit. After a short period of self-regulation, the scores rise up to a new maximum smaller than the previous one when the remaining UAVs reach the new optimal positions. The proposed PSR-APC and the passive reaction approach differ around the timing of UAV quit. The passive reaction approach has no reaction before UAV quit and thus experiences dramatic drop in US scores. On the contrary, the PSR-APC approach monitors the UAV battery status and starts moving the UAVs one or two epochs before the UAV quit. Although the epoch US scores may drop early due to pre-movements, they will not drop that low when the UAV quit as those under the passive reaction approach. Besides, the transition process will be completed earlier. As a result, the accumulated US scores during the transition are higher than those under the passive reaction approach, as shown in Fig. \ref{fig.Gain_Quit}. 
However, proactive movement is not always considerably beneficial since the gain depends on specific user distribution and user-to-UAV ratio. When $N_{UAV}=4$, the gain is marginal. The reason is that before one UAV quits, the 4 UAVs are separately positioned over 4 hot spots far away from each other. When one UAV quits, at least one UAV needs to move a long way to the next optimal position, along which the epoch US score even drops lower. In such a situation, the gain of pre-movements are significantly diluted by the long transition period.

\subsubsection{\textbf{Case of UAV Join-in}}
\begin{figure}[!ht]
\begin{tabular}{l}
\subfigure[$N_{UAV}=5$] {\includegraphics[width=2.8in]{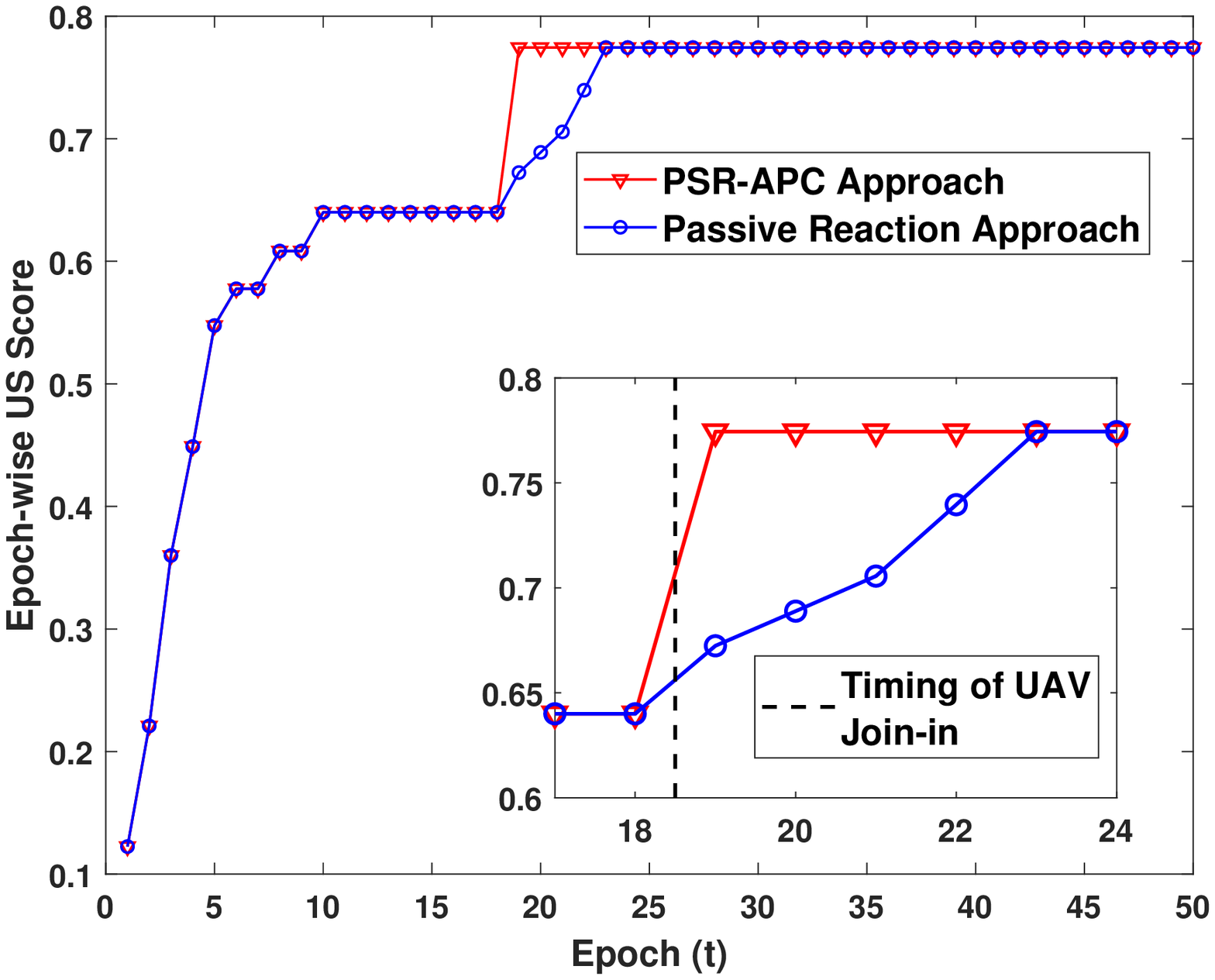}\label{Fig.N5Join}} \\
\subfigure[$N_{UAV}=4$] {\includegraphics[width=2.8in]{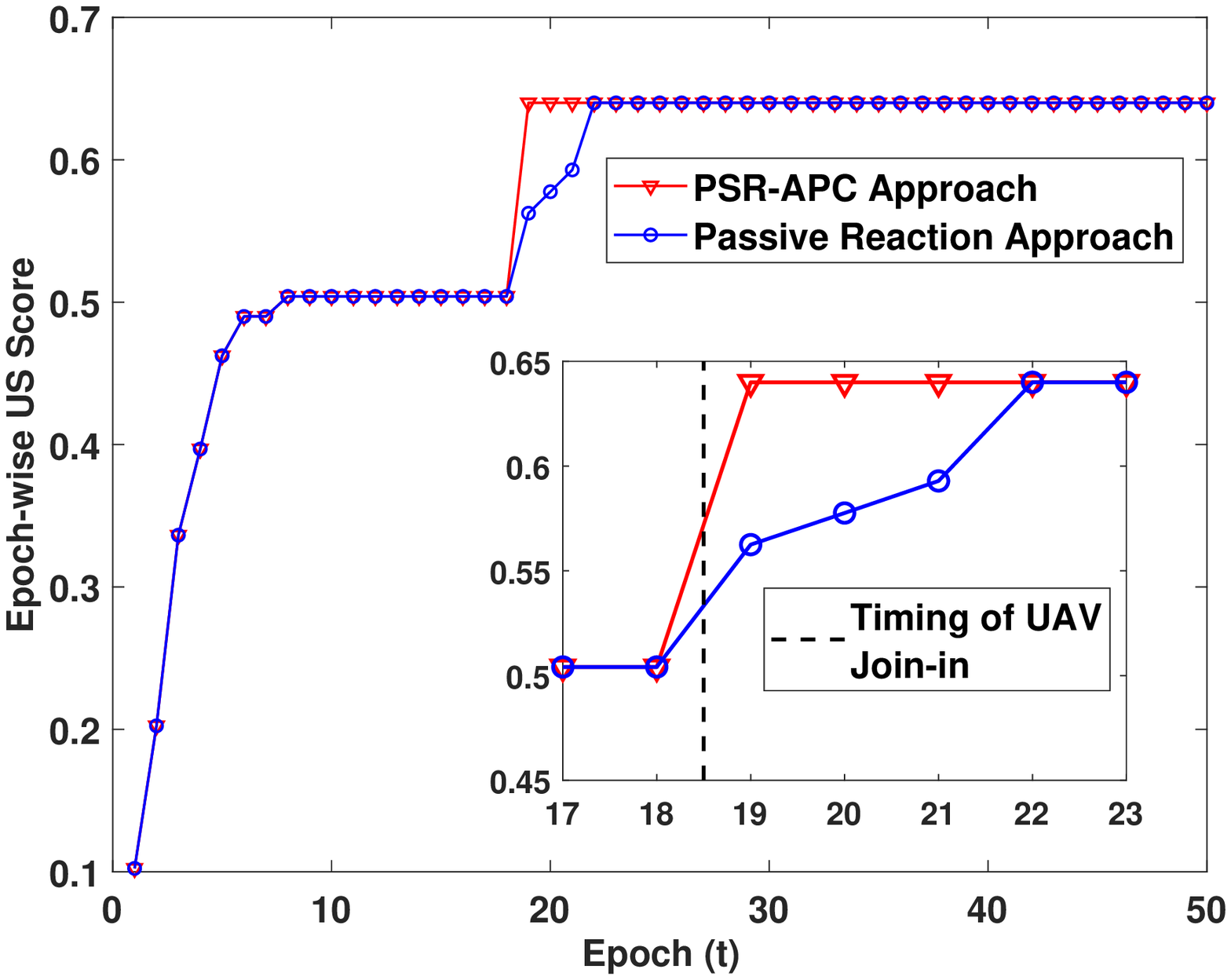}\label{Fig.N4Join}} \\
\subfigure[$N_{UAV}=3$] {\includegraphics[width=2.8in]{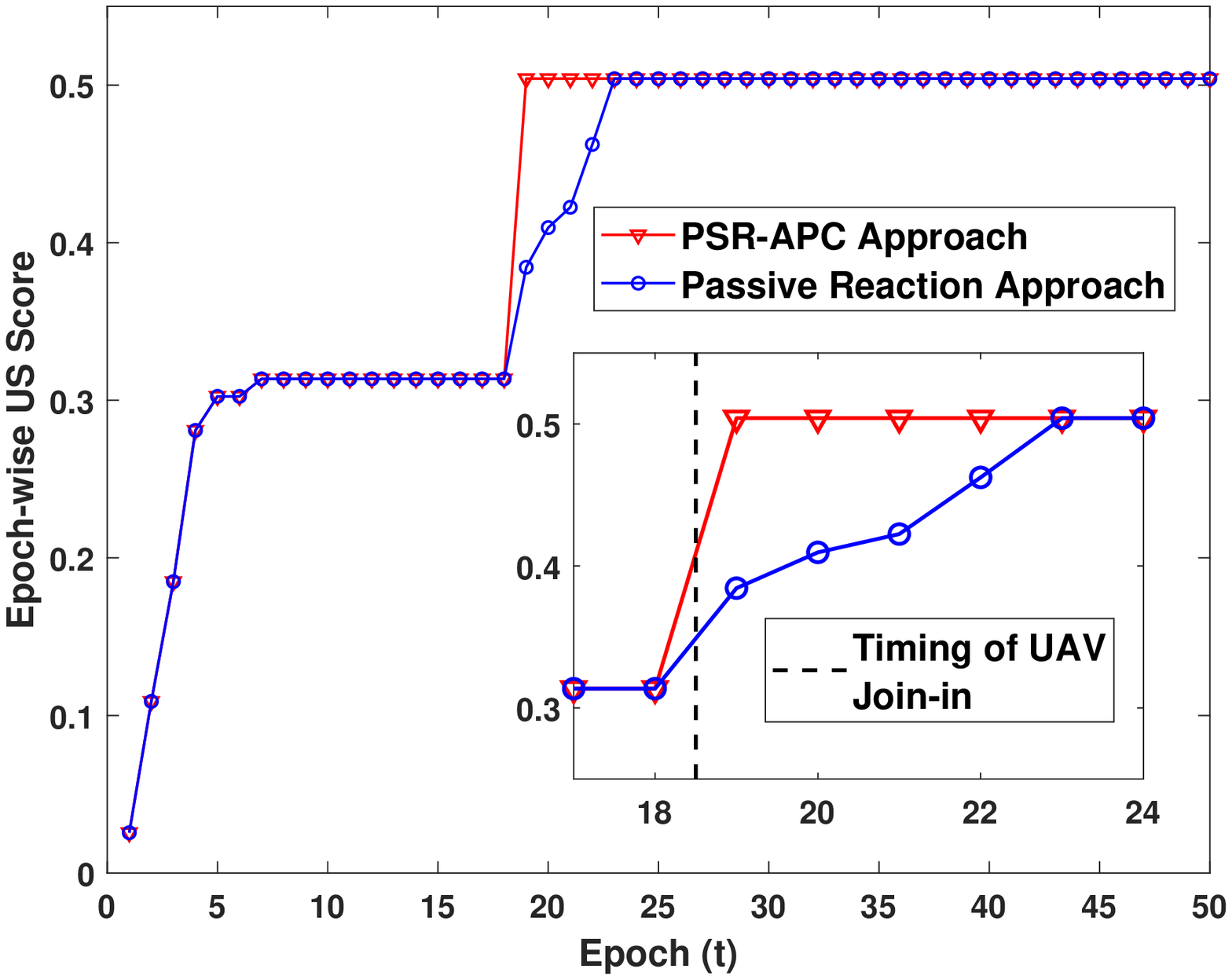}\label{Fig.N3Join}}
\end{tabular}
\caption{\small{Case of UAV join-in: Epoch-wise reward comparison between the PSR-APC approach and the passive reaction approach.}}\label{fig.Epoch_Join}
\end{figure}
\begin{figure}[!ht]
	\centering
	\includegraphics[width=3.3in]{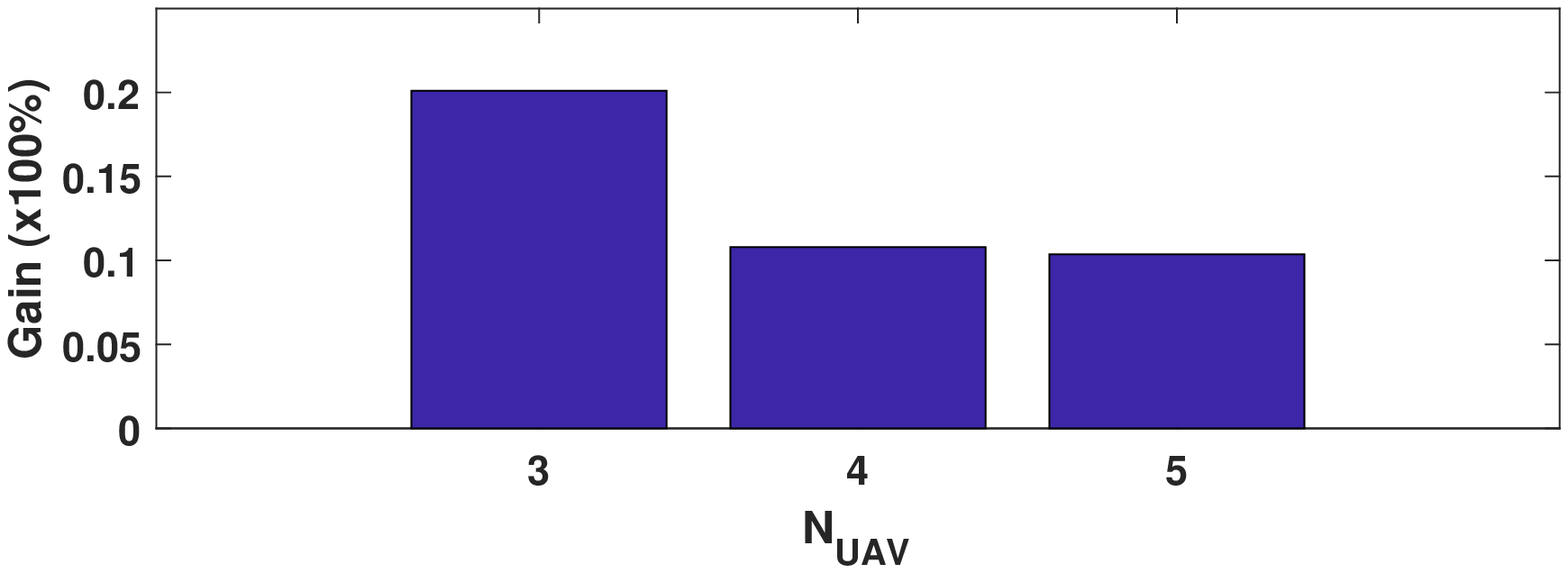}
	\caption{\small{Case of UAV join-in: Gain ($\%$) of PSR-APC over the passive reaction approach in accumulated US scores during transition to the new optimal UAV positions.}} \label{fig.Gain_Join}
\end{figure}

The case of UAV join-in is then simulated with different $N_{UAV}$. The epoch-wise US scores under both the PSR-APC approach and the passive reaction approach are presented in Fig. \ref{fig.Epoch_Join}. There are initially $N_{UAV}$ UAVs that start off at the unit circular positions, and then reach the optimal positions. A joining UAV starts elevating from the ground in the center (5,5) at epoch 11, and reaches the serving altitude (i.e., formally join the network) at epoch 19. The passive reaction approach in this case relocates UAVs only after the joining UAV joins the network right above (5,5), while the PSR-APC approach starts tuning the horizontal positions of the joining UAV right after it starts off. This ensures that when the joining UAV reaches the serving altitude, all the UAVs are already near the new optimal positions. This is confirmed by the curves in Fig. \ref{fig.Epoch_Join}. It can be observed that under the PSR-APC approach, all the UAVs are at the new optimal positions in the very first epoch after the new UAV joins in, while it takes the passive approach couple of epochs to dispatch the UAVs to the new optimal positions. In addition, there is no pre-movement of the existing UAVs when the new UAV is about to join. This is because the new optimal positions of the existing UAVs are within 1 epoch reach to the previous optimal positions in our user distribution settings. The gain in accumulated US score during the transition period introduced by the PSR-APC approach is shown in Fig. \ref{fig.Gain_Join}, achieving at least 10$\%$ for all simulated $N_{UAV}$.

\subsubsection{\textbf{Case of UAV and User Dynamics}}
\begin{figure}[!ht]
	\centering
	\includegraphics[width=3.0in]{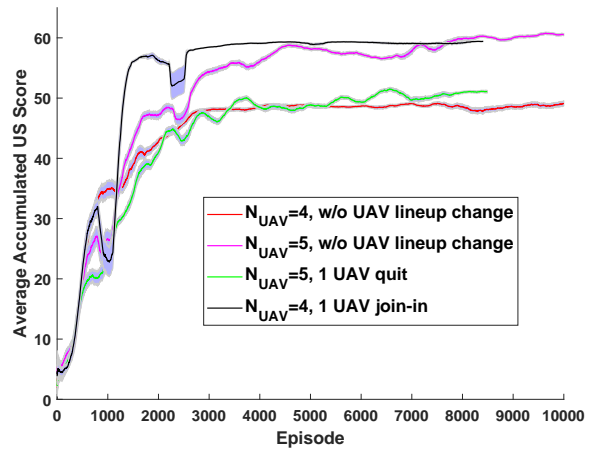}
	\caption{Convergence with 95$\%$ credit interval with user dynamics.} \label{fig.User_Converge}
\end{figure}

\begin{figure*}[!ht]
\centering
\begin{tabular}{l}
\subfigure[No UAV lineup change] {\includegraphics[width=2.1in]{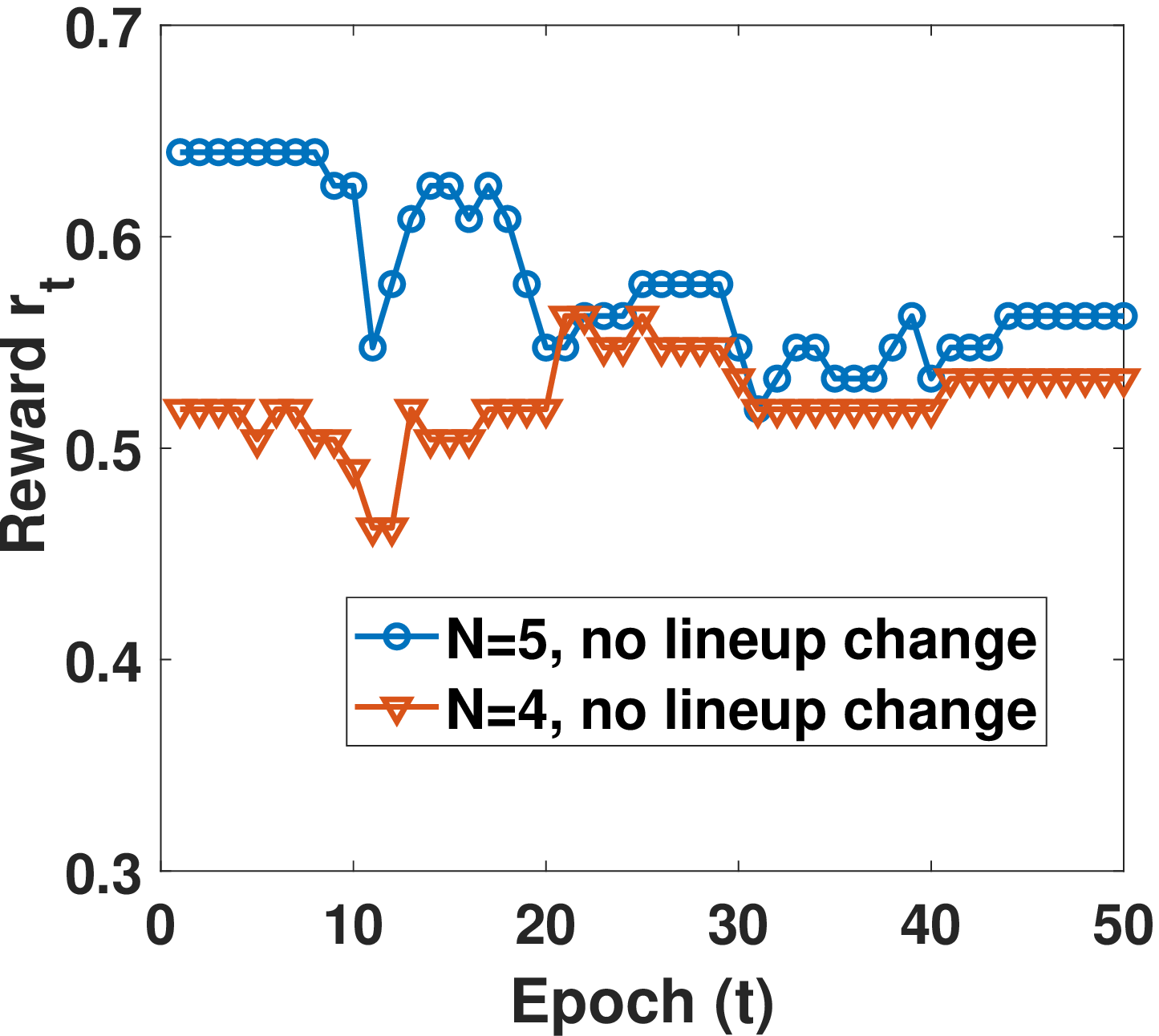}\label{Fig.Epoch1}}\;\;
\subfigure[$N_{UAV}=5$ with one UAV quit] {\includegraphics[width=2.0in]{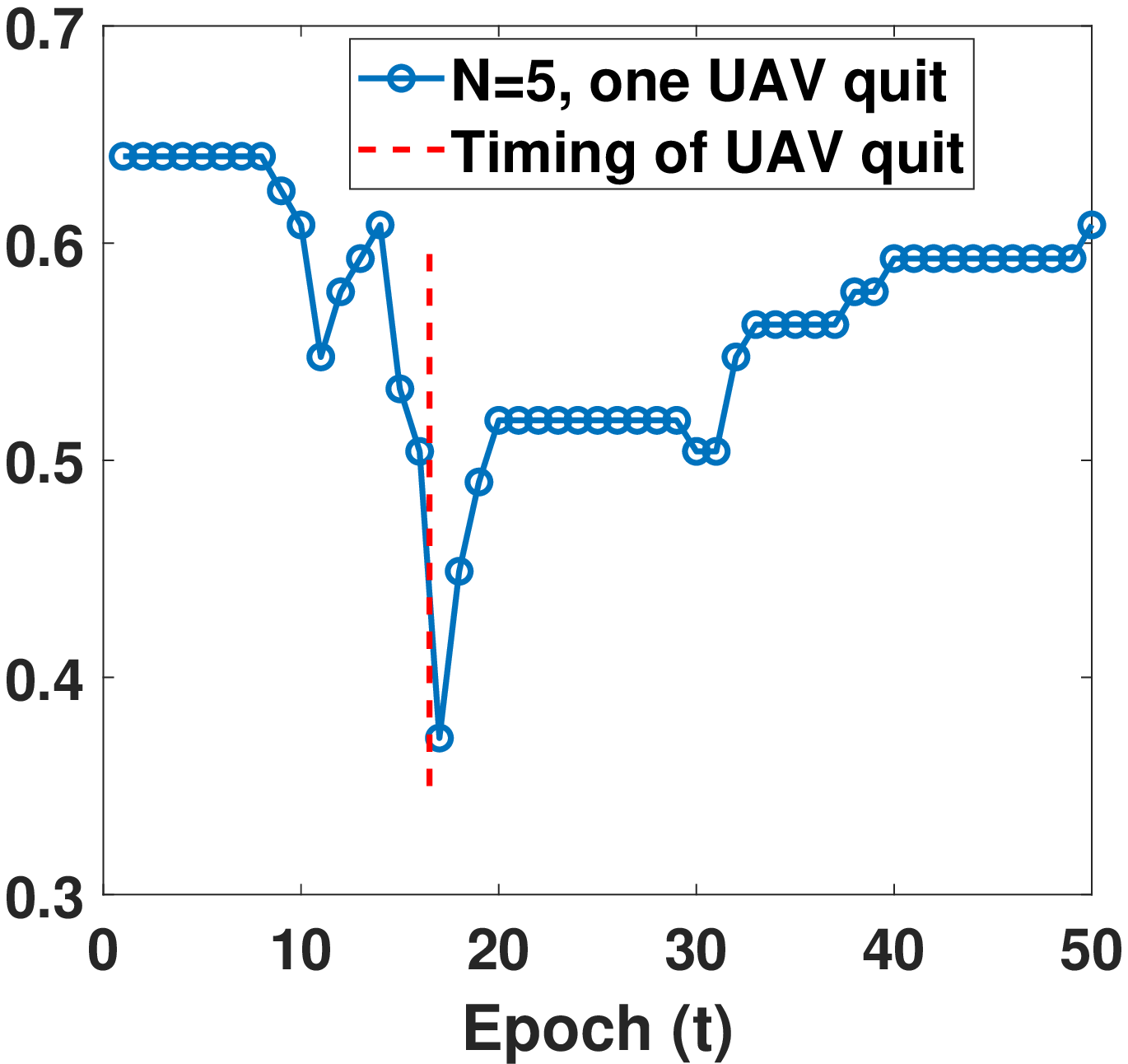}\label{Fig.Epoch2}}\;\;
\subfigure[$N_{UAV}=4$ with one UAV join-in] {\includegraphics[width=2.0in,height=1.9in]{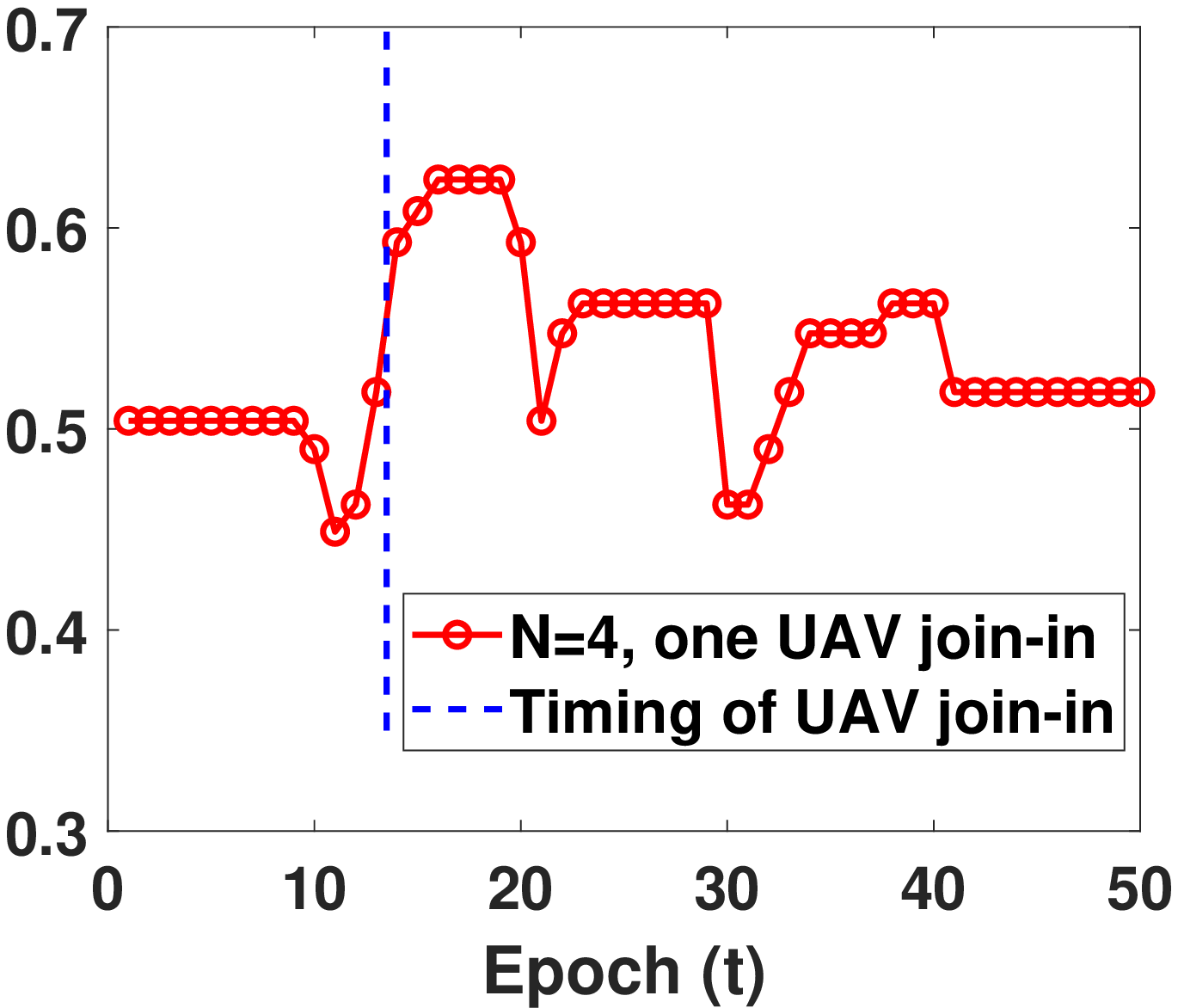}\label{Fig.Epoch3}}
\end{tabular}
\caption{Epoch-wise reward of different situations with dynamic user distributions.}\label{fig.Epochs}
\end{figure*}

At last, the case with dynamic user distribution is simulated. The disperse-gather-disperse procedure shown in Fig. \ref{fig.distribution} is employed. We divide the 100 epochs evenly into 10 segments. The centers of the hot spots are updated (i.e., user distribution is updated) at the beginning of each segment. The order of update is snapshot 1$\rightarrow$2$\rightarrow$3$\rightarrow$4$\rightarrow$4$\rightarrow$4$\rightarrow$4$\rightarrow$3$\rightarrow$2$\rightarrow$1. The $N_{UAV}$ UAVs start off initially from the optimal positions of snapshot 1, and move accordingly while the user distribution changes. The convergence of 4 situations is shown in Fig. \ref{fig.User_Converge}: 4 UAVs and 5 UAVs with no UAV quit or join-in, 5 UAVs with 1 UAV quit, and 4 UAVs with 1 UAV join-in. In addition, the epoch-wise reward of each situation is presented in Fig. \ref{fig.Epochs}. It can be observed that the epoch-wise rewards are relatively steady within each time segment (where the hotspot centers remain still), but experience considerable changes when crossing the time segments. As shown in Subfigures \ref{fig.Epochs}(b)(c), our proposed approach can also handle the change in UAV lineup with time-varying user distribution, by getting the UAVs to the new optimal positions soon after the change.

\begin{figure*}[!ht]
\centering
\begin{tabular}{l}
\subfigure[$N_{UAV}=4$ w/o UAV lineup change] {\includegraphics[width=2.3in]{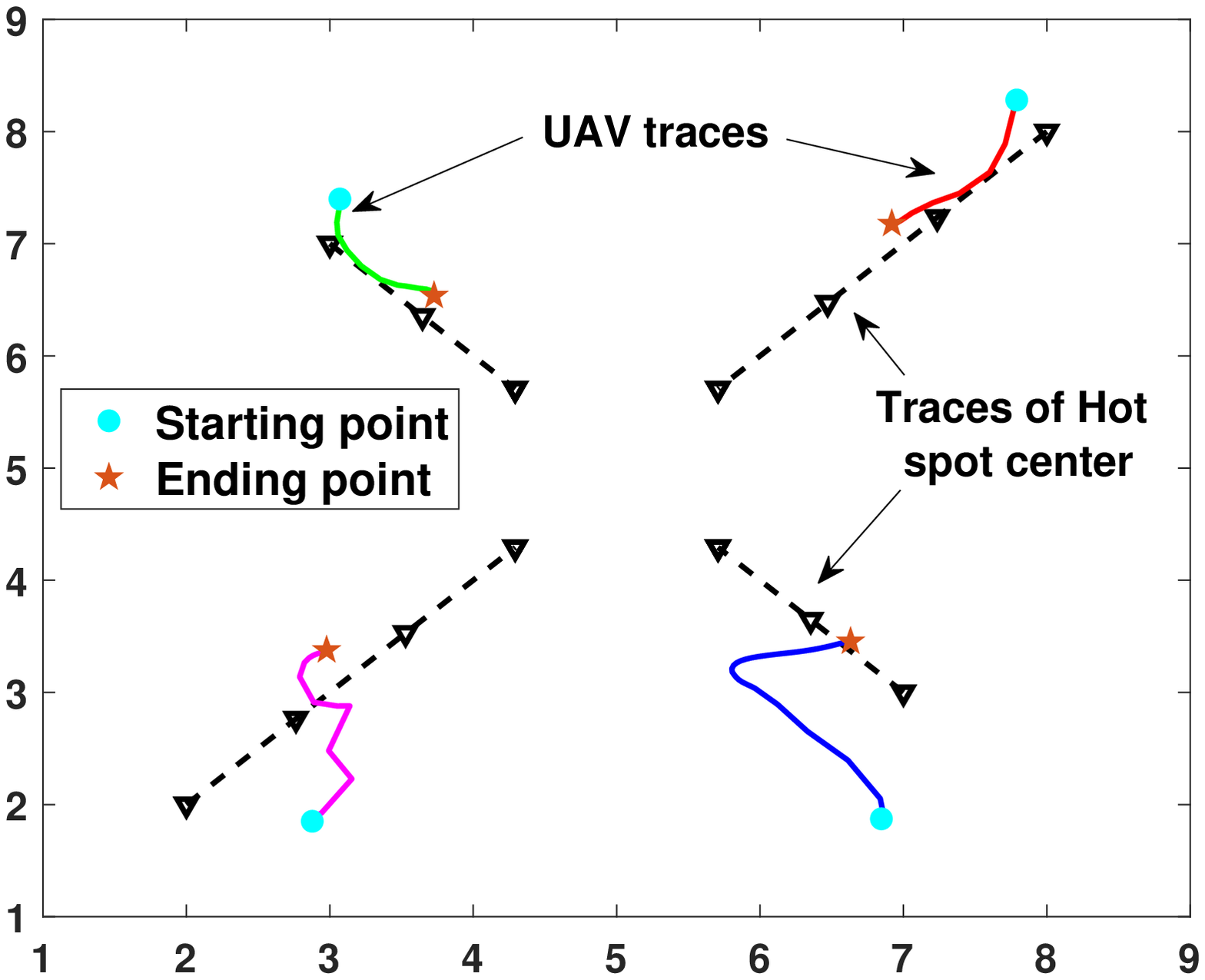}\label{Fig.Trace1}}\;\;\;\;\;\;\;\;\; 
\subfigure[$N_{UAV}=5$ w/o UAV lineup change] {\includegraphics[width=2.3in,height=1.85in]{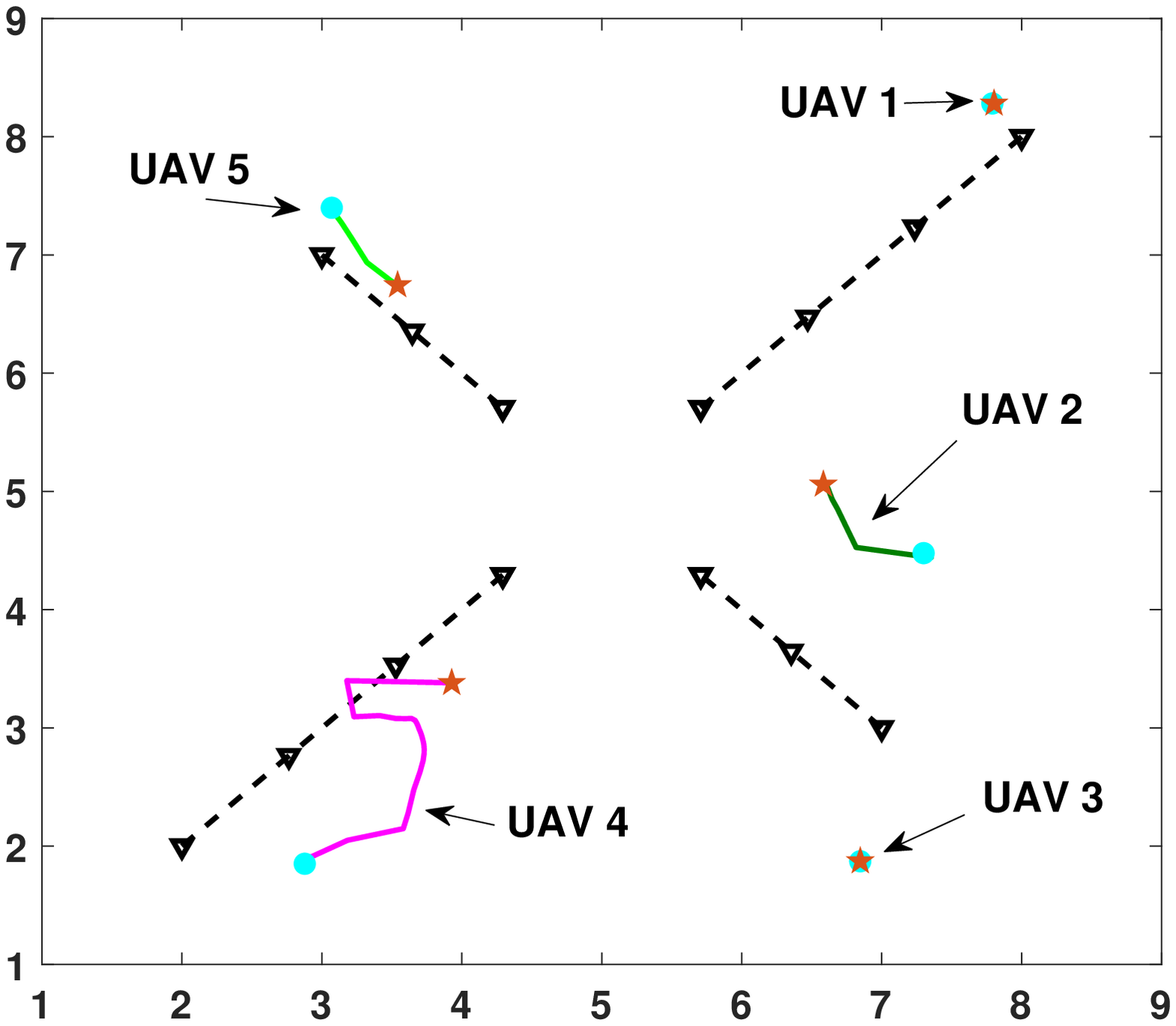}\label{Fig.Trace2}} \\
\subfigure[$N_{UAV}=5$ with one UAV quit] {\includegraphics[width=2.3in]{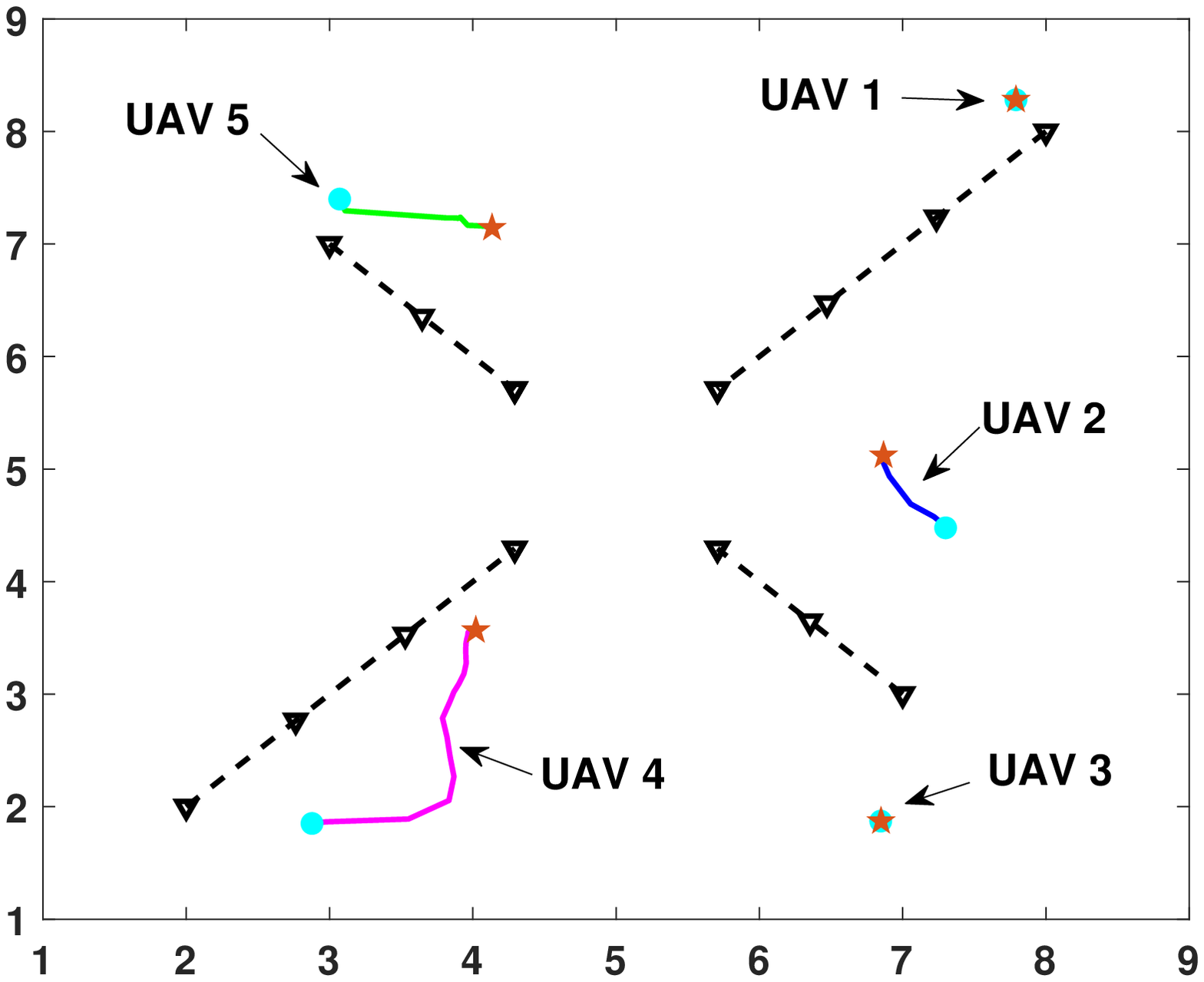}\label{Fig.Trace3}}\;\;\;\;\;\;\;\;\; 
\subfigure[$N_{UAV}=4$ with one UAV join-in] {\includegraphics[width=2.3in,height=1.85in]{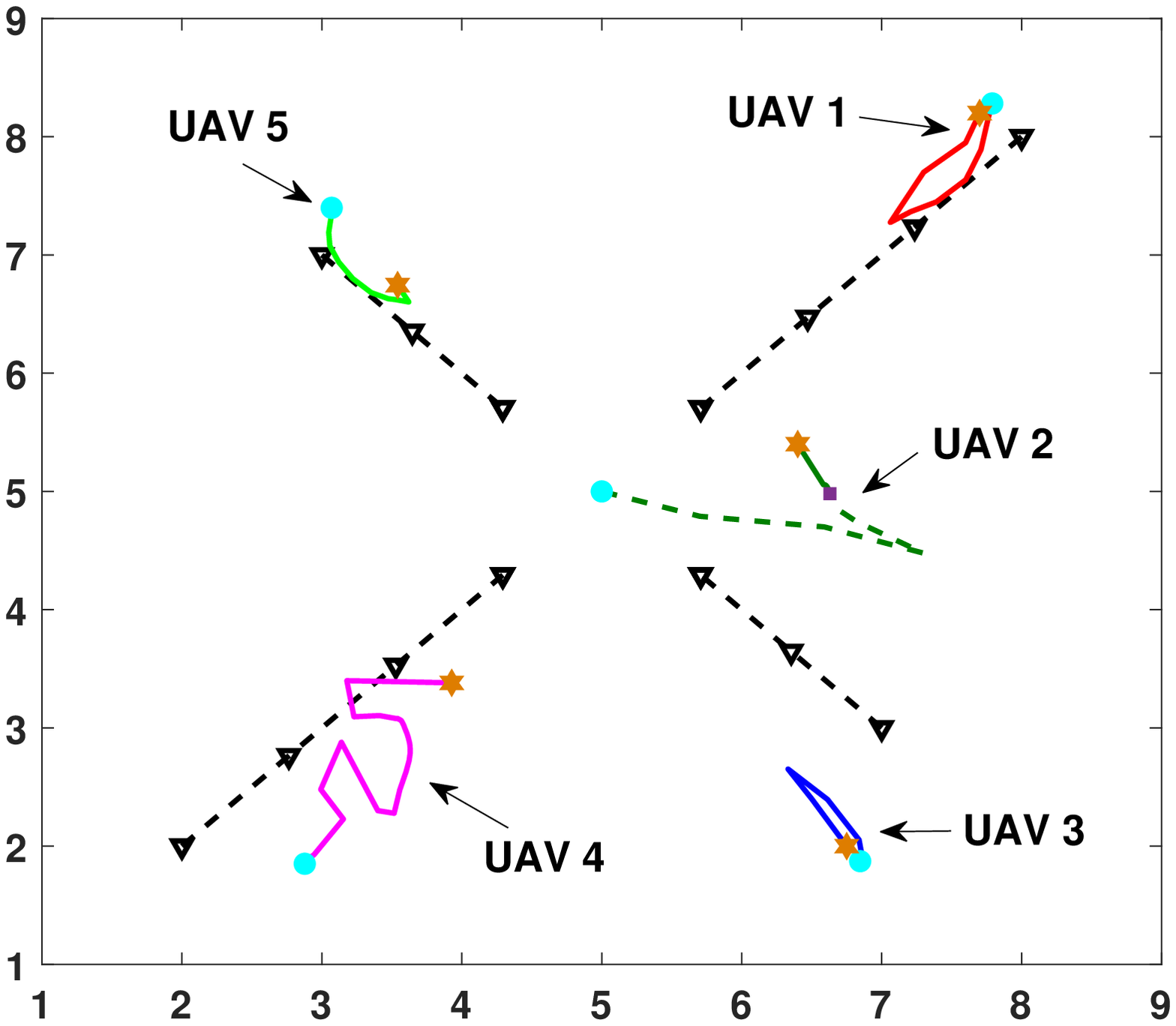}\label{Fig.Trace4}}
\end{tabular}
\caption{Optimal UAV trajectories with user distribution dynamics. The x and y axis represent the horizontal coordinates.}\label{fig.Traces}
\end{figure*}

The UAV trajectories are also demonstrated in Fig. \ref{fig.Traces}. As the disperse-gather procedure is just the opposite mirror of the gather-disperse procedure, we only present the first 50 epochs. In all 4 subfigures, the black dashed lines represent the traces of the hotspot centers, moving from corners towards the center of the target region, and stops 1 unit away from the center. In Subfigure \ref{Fig.Trace1}, as the hotspots move, the 4 UAVs proactively move from the initial positions (solid aqua circles) towards the region center to cover as many users as possible. But instead of exactly following the hotspot centers, the UAVs stop farther from the region center (solid brown pentagrams). This is because $i$) the ground coverage radius of each UAV is larger than 1 unit, and $ii$) coverage overlapping of UAVs will lead to significant intercell interference and further affect the user QoS. Situations are different when there are 5 UAVs, as shown in Subfigure \ref{Fig.Trace2}. Interestingly, while hotspots move, UAV 1 and UAV 3 almost stay still because of the existence of UAV 2. As UAV 2 moves towards the region center, it is able to cover most of the user flows from hotspots in the top and bottom right corners, so that UAV 1 and UAV 3 can stay put to cover more uniformly distributed users. Accordingly, the trace of UAV 4 leans a little towards the center to help cover the center users, and the trace of UAV 5 backs off a little to reduce overlapping. 

Subfigure \ref{Fig.Trace3} shows the UAV traces when UAV 1 quits the network around epoch 16. It can be observed that after UAV 1 quits, the traces of UAV 2 and 5 turn more towards the hotspot in the top right corner to cover more users. UAV 4 goes deeper towards the region center to cover the users missed by UAV 2 and 5 while UAV 3 stays put. Subfigure \ref{Fig.Trace4} shows the UAV traces when UAV 2 starts off at the region center and formally joins the network at epoch 14. The dashed part of UAV 2 represents horizontal trace before it reaches the serving altitude. It can be observed that before UAV 2 joins the network, the 4 existing UAVs move very similarly to Subfigure \ref{Fig.Trace1}. After UAV 2 joins, UAV 1 and 3 gradually back to the proximity of their original starting points, while UAV 4 and 5 start to move like subfigure \ref{Fig.Trace2} where there are 5 existing UAVs.
\section{Conclusions}\label{sec.Conclusion}
In this chapter, an RL-based regulation strategy of a UAV communication network has been investigated with dynamic UAV lineup and user distribution. The learning approach, i.e, PSR-APC, has been designed to responsively control the UAV trajectories when the group of serving UAVs or the user distribution change within a considered time horizon. Simulation results have demonstrated that compared to the passive reaction approach, the proposed approach can achieve up to 20$\%$ higher accumulated US scores during the transition process. In addition, when the user distribution is dynamically changing, the proposed approach has been shown to be able to capture the dynamics and move UAVs accordingly.

\bibliographystyle{IEEEtran}
\bibliography{reference}

\begin{thebibliography}{10}
\providecommand{\url}[1]{#1}
\csname url@samestyle\endcsname
\providecommand{\newblock}{\relax}
\providecommand{\bibinfo}[2]{#2}
\providecommand{\BIBentrySTDinterwordspacing}{\spaceskip=0pt\relax}
\providecommand{\BIBentryALTinterwordstretchfactor}{4}
\providecommand{\BIBentryALTinterwordspacing}{\spaceskip=\fontdimen2\font plus
\BIBentryALTinterwordstretchfactor\fontdimen3\font minus
  \fontdimen4\font\relax}
\providecommand{\BIBforeignlanguage}[2]{{%
\expandafter\ifx\csname l@#1\endcsname\relax
\typeout{** WARNING: IEEEtran.bst: No hyphenation pattern has been}%
\typeout{** loaded for the language `#1'. Using the pattern for}%
\typeout{** the default language instead.}%
\else
\language=\csname l@#1\endcsname
\fi
#2}}
\providecommand{\BIBdecl}{\relax}
\BIBdecl

\bibitem{zhang2019iot}
Q.~Zhang, M.~Jiang, Z.~Feng, W.~Li, W.~Zhang, and M.~Pan,
  ``\uppercase{I}o\uppercase{T} enabled \uppercase{UAV}: Network architecture
  and routing algorithm,'' \emph{IEEE Internet of Things Journal}, vol.~6,
  no.~2, pp. 3727--3742, 2019.

\bibitem{zeng2016wireless}
Y.~Zeng, R.~Zhang, and T.~J. Lim, ``Wireless communications with unmanned
  aerial vehicles: Opportunities and challenges,'' \emph{IEEE Communications
  Magazine}, vol.~54, no.~5, pp. 36--42, 2016.

\bibitem{market}
``Unmanned aerial vehicle (\uppercase{UAV}) market,'' Available at
  \url{https://www.marketsandmarkets.com/Market-Reports/unmanned-aerial-vehicles-uav-market-662.html},
  Octorber 2019.

\bibitem{li2020energy}
M.~Li, N.~Cheng, J.~Gao, Y.~Wang, L.~Zhao, and X.~Shen, ``Energy-efficient
  \uppercase{UAV}-assisted mobile edge computing: Resource allocation and
  trajectory optimization,'' \emph{IEEE Transactions on Vehicular Technology},
  vol.~69, no.~3, pp. 3424--3438, 2020.

\bibitem{shi2020mean}
D.~Shi, H.~Gao, L.~Wang, M.~Pan, Z.~Han, and H.~V. Poor, ``Mean field game
  guided deep reinforcement learning for task placement in cooperative
  multi-access edge computing,'' \emph{IEEE Internet of Things Journal}, 2020.

\bibitem{motlagh2017uav}
N.~H. Motlagh, M.~Bagaa, and T.~Taleb, ``\uppercase{UAV}-based
  \uppercase{I}o\uppercase{T} platform: A crowd surveillance use case,''
  \emph{IEEE Communications Magazine}, vol.~55, no.~2, pp. 128--134, 2017.

\bibitem{zhao2019disaster}
N.~Zhao, W.~Lu, M.~Sheng, Y.~Chen, J.~Tang, F.~R. Yu, and K.-K. Wong,
  ``\uppercase{UAV}-assisted emergency networks in disasters,'' \emph{IEEE
  Wireless Communications}, vol.~26, no.~1, pp. 45--51, 2019.

\bibitem{chen2017caching}
M.~Chen, M.~Mozaffari, W.~Saad, C.~Yin, M.~Debbah, and C.~S. Hong, ``Caching in
  the sky: Proactive deployment of cache-enabled unmanned aerial vehicles for
  optimized quality-of-experience,'' \emph{IEEE Journal on Selected Areas in
  Communications}, vol.~35, no.~5, pp. 1046--1061, 2017.

\bibitem{nasir2019uav}
A.~A. Nasir, H.~D. Tuan, T.~Q. Duong, and H.~V. Poor, ``\uppercase{UAV}-enabled
  communication using noma,'' \emph{IEEE Transactions on Communications},
  vol.~67, no.~7, pp. 5126--5138, 2019.

\bibitem{mozaffari2019tutorial}
M.~Mozaffari, W.~Saad, M.~Bennis, Y.-H. Nam, and M.~Debbah, ``A tutorial on
  \uppercase{UAV}s for wireless networks: Applications, challenges, and open
  problems,'' \emph{IEEE Communications Surveys \& Tutorials}, vol.~21, no.~3,
  pp. 2334--2360, 2019.

\bibitem{wu2019fundamental}
Q.~Wu, L.~Liu, and R.~Zhang, ``Fundamental trade-offs in communication and
  trajectory design for \uppercase{UAV}-enabled wireless network,'' \emph{IEEE
  Wireless Communications}, vol.~26, no.~1, pp. 36--44, 2019.

\bibitem{zeng2018trajectory}
Y.~Zeng, X.~Xu, and R.~Zhang, ``Trajectory design for completion time
  minimization in \uppercase{UAV}-enabled multicasting,'' \emph{IEEE
  Transactions on Wireless Communications}, vol.~17, no.~4, pp. 2233--2246,
  2018.

\bibitem{wu2018common}
Q.~Wu and R.~Zhang, ``Common throughput maximization in \uppercase{UAV}-enabled
  \uppercase{OFDMA} systems with delay consideration,'' \emph{IEEE Transactions
  on Communications}, vol.~66, no.~12, pp. 6614--6627, 2018.

\bibitem{zeng2019energy}
Y.~Zeng, J.~Xu, and R.~Zhang, ``Energy minimization for wireless communication
  with rotary-wing \uppercase{UAV},'' \emph{IEEE Transactions on Wireless
  Communications}, vol.~18, no.~4, pp. 2329--2345, 2019.

\bibitem{guo2019uav}
H.~Guo and J.~Liu, ``\uppercase{UAV}-enhanced intelligent offloading for
  internet of things at the edge,'' \emph{IEEE Transactions on Industrial
  Informatics}, vol.~16, no.~4, pp. 2737--2746, 2019.

\bibitem{mozaffari2018beyond}
M.~Mozaffari, A.~T.~Z. Kasgari, W.~Saad, M.~Bennis, and M.~Debbah, ``Beyond
  \uppercase{5G} with \uppercase{UAV}s: Foundations of a 3\uppercase{D}
  wireless cellular network,'' \emph{IEEE Transactions on Wireless
  Communications}, vol.~18, no.~1, pp. 357--372, 2018.

\bibitem{wu2018joint}
Q.~Wu, Y.~Zeng, and R.~Zhang, ``Joint trajectory and communication design for
  multi-\uppercase{UAV} enabled wireless networks,'' \emph{IEEE Transactions on
  Wireless Communications}, vol.~17, no.~3, pp. 2109--2121, 2018.

\bibitem{mozaffari2017mobile}
M.~Mozaffari, W.~Saad, M.~Bennis, and M.~Debbah, ``Mobile unmanned aerial
  vehicles (\uppercase{UAV}s) for energy-efficient internet of things
  communications,'' \emph{IEEE Transactions on Wireless Communications},
  vol.~16, no.~11, pp. 7574--7589, 2017.

\bibitem{yang2019energy}
Z.~Yang, C.~Pan, K.~Wang, and M.~Shikh-Bahaei, ``Energy efficient resource
  allocation in \uppercase{UAV}-enabled mobile edge computing networks,''
  \emph{IEEE Transactions on Wireless Communications}, vol.~18, no.~9, pp.
  4576--4589, 2019.

\bibitem{alpaydin2020introduction}
E.~Alpaydin, \emph{Introduction to machine learning}.\hskip 1em plus 0.5em
  minus 0.4em\relax MIT press, 2020.

\bibitem{RL}
R.~S. Sutton and A.~G. Barto, \emph{Reinforcement Learning: An Introduction,
  2nd Edition}.\hskip 1em plus 0.5em minus 0.4em\relax Bradford Books, 2018.

\bibitem{luong2019applications}
N.~C. Luong, D.~T. Hoang, S.~Gong, D.~Niyato, P.~Wang, Y.-C. Liang, and D.~I.
  Kim, ``Applications of deep reinforcement learning in communications and
  networking: A survey,'' \emph{IEEE Communications Surveys \& Tutorials},
  vol.~21, no.~4, pp. 3133--3174, 2019.

\bibitem{klaine2018distributed}
P.~V. Klaine, J.~P. Nadas, R.~D. Souza, and M.~A. Imran, ``Distributed drone
  base station positioning for emergency cellular networks using reinforcement
  learning,'' \emph{Cognitive computation}, vol.~10, no.~5, pp. 790--804, 2018.

\bibitem{cui2019multi}
J.~Cui, Y.~Liu, and A.~Nallanathan, ``Multi-agent reinforcement learning-based
  resource allocation for \uppercase{UAV} networks,'' \emph{IEEE Transactions
  on Wireless Communications}, vol.~19, no.~2, pp. 729--743, 2019.

\bibitem{hu2020reinforcement}
J.~Hu, H.~Zhang, L.~Song, Z.~Han, and H.~V. Poor, ``Reinforcement learning for
  a cellular internet of \uppercase{UAV}s: protocol design, trajectory control,
  and resource management,'' \emph{IEEE Wireless Communications}, vol.~27,
  no.~1, pp. 116--123, 2020.

\bibitem{liu2019trajectory}
X.~Liu, Y.~Liu, Y.~Chen, and L.~Hanzo, ``Trajectory design and power control
  for multi-\uppercase{UAV} assisted wireless networks: A machine learning
  approach,'' \emph{IEEE Transactions on Vehicular Technology}, vol.~68, no.~8,
  pp. 7957--7969, 2019.

\bibitem{hu2020distributed}
Y.~Hu, M.~Chen, W.~Saad, H.~V. Poor, and S.~Cui, ``Distributed multi-agent meta
  learning for trajectory design in wireless drone networks,'' \emph{arXiv
  preprint arXiv:2012.03158}, 2020.

\bibitem{singh2018distributed}
S.~Singh, A.~Kumbhar, I.~G{\"u}ven{\c{c}}, and M.~L. Sichitiu, ``Distributed
  approaches for inter-cell interference coordination in \uppercase{UAV}-based
  \uppercase{lte-a}dvanced \uppercase{H}et\uppercase{N}ets,'' in \emph{2018
  IEEE 88th Vehicular Technology Conference (VTC-Fall)}.\hskip 1em plus 0.5em
  minus 0.4em\relax IEEE, 2018, pp. 1--6.

\bibitem{challita2019interference}
U.~Challita, W.~Saad, and C.~Bettstetter, ``Interference management for
  cellular-connected \uppercase{UAV}s: A deep reinforcement learning
  approach,'' \emph{IEEE Transactions on Wireless Communications}, vol.~18,
  no.~4, pp. 2125--2140, 2019.

\bibitem{tang2020deep}
F.~Tang, Y.~Zhou, and N.~Kato, ``Deep reinforcement learning for dynamic
  uplink/downlink resource allocation in high mobility \uppercase{5G}
  \uppercase{H}et\uppercase{N}et,'' \emph{IEEE Journal on Selected Areas in
  Communications}, 2020.

\bibitem{cheng2019space}
N.~Cheng, F.~Lyu, W.~Quan, C.~Zhou, H.~He, W.~Shi, and X.~Shen,
  ``Space/aerial-assisted computing offloading for \uppercase{I}o\uppercase{T}
  applications: A learning-based approach,'' \emph{IEEE Journal on Selected
  Areas in Communications}, vol.~37, no.~5, pp. 1117--1129, 2019.

\bibitem{liu2019optimized}
X.~Liu, M.~Chen, and C.~Yin, ``Optimized trajectory design in \uppercase{UAV}
  based cellular networks for 3\uppercase{D} users: A double
  \uppercase{Q}-learning approach,'' \emph{arXiv preprint arXiv:1902.06610},
  2019.

\bibitem{liu2018energy}
C.~H. Liu, Z.~Chen, J.~Tang, J.~Xu, and C.~Piao, ``Energy-efficient
  \uppercase{UAV} control for effective and fair communication coverage: A deep
  reinforcement learning approach,'' \emph{IEEE Journal on Selected Areas in
  Communications}, vol.~36, no.~9, pp. 2059--2070, 2018.

\bibitem{khairy2020constrained}
S.~Khairy, P.~Balaprakash, L.~X. Cai, and Y.~Cheng, ``Constrained deep
  reinforcement learning for energy sustainable multi-\uppercase{UAV} based
  random access \uppercase{I}o\uppercase{T} networks with \uppercase{NOMA},''
  \emph{arXiv preprint arXiv:2002.00073}, 2020.

\bibitem{huang2020online}
Y.~Huang, X.~Mo, J.~Xu, L.~Qiu, and Y.~Zeng, ``Online maneuver design for
  \uppercase{UAV}-enabled noma systems via reinforcement learning,'' in
  \emph{2020 IEEE Wireless Communications and Networking Conference
  (WCNC)}.\hskip 1em plus 0.5em minus 0.4em\relax IEEE, 2020, pp. 1--6.

\bibitem{lillicrap2015continuous}
T.~P. Lillicrap, J.~J. Hunt, A.~Pritzel, N.~Heess, T.~Erez, Y.~Tassa,
  D.~Silver, and D.~Wierstra, ``Continuous control with deep reinforcement
  learning,'' \emph{arXiv preprint arXiv:1509.02971}, 2015.

\bibitem{pham2018cooperative}
H.~X. Pham, H.~M. La, D.~Feil-Seifer, and A.~Nefian, ``Cooperative and
  distributed reinforcement learning of drones for field coverage,''
  \emph{arXiv preprint arXiv:1803.07250}, 2018.

\bibitem{chen2020mean}
D.~Chen, Q.~Qi, Z.~Zhuang, J.~Wang, J.~Liao, and Z.~Han, ``Mean field deep
  reinforcement learning for fair and efficient uav control,'' \emph{IEEE
  Internet of Things Journal}, vol.~8, no.~2, pp. 813--828, 2020.

\bibitem{al2014modeling}
A.~Al-Hourani, S.~Kandeepan, and A.~Jamalipour, ``Modeling air-to-ground path
  loss for low altitude platforms in urban environments,'' in \emph{2014 IEEE
  global communications conference}.\hskip 1em plus 0.5em minus 0.4em\relax
  IEEE, 2014, pp. 2898--2904.

\bibitem{seddon2011basic}
J.~M. Seddon and S.~Newman, \emph{Basic helicopter aerodynamics}.\hskip 1em
  plus 0.5em minus 0.4em\relax John Wiley \& Sons, 2011, vol.~40.

\bibitem{han2020}
M.~Han, S.~Khairy, L.~X. Cai, Y.~Cheng, and R.~Zhang, ``Reinforcement learning
  for efficient and fair coexistence between \uppercase{LTE-LAA} and
  \uppercase{W}i-\uppercase{F}i,'' \emph{IEEE Transactions on Vehicular
  Technology}, to appear.

\bibitem{shi2019deep}
D.~Shi, J.~Ding, S.~M. Errapotu, H.~Yue, W.~Xu, X.~Zhou, and M.~Pan, ``Deep
  \uppercase{Q}-network-based route scheduling for \uppercase{TNC} vehicles
  with passengers’ location differential privacy,'' \emph{IEEE Internet of
  Things Journal}, vol.~6, no.~5, pp. 7681--7692, 2019.

\bibitem{hester2018deep}
T.~Hester, M.~Vecerik, O.~Pietquin, M.~Lanctot, T.~Schaul, B.~Piot, D.~Horgan,
  J.~Quan, A.~Sendonaris, I.~Osband \emph{et~al.}, ``Deep
  \uppercase{q}-learning from demonstrations,'' in \emph{Thirty-Second AAAI
  Conference on Artificial Intelligence}, 2018.

\bibitem{mnih2016asynchronous}
V.~Mnih, A.~P. Badia, M.~Mirza, A.~Graves, T.~Lillicrap, T.~Harley, D.~Silver,
  and K.~Kavukcuoglu, ``Asynchronous methods for deep reinforcement learning,''
  in \emph{International conference on machine learning}, 2016, pp. 1928--1937.

\bibitem{zeng2017energy}
Y.~Zeng and R.~Zhang, ``Energy-efficient \uppercase{UAV} communication with
  trajectory optimization,'' \emph{IEEE Transactions on Wireless
  Communications}, vol.~16, no.~6, pp. 3747--3760, 2017.

\end{thebibliography}

\end{document}